\documentclass[notitlepage, superscriptaddress, 11pt, single-spaced,floatfix,nofootinbib]{revtex4-2}
\usepackage[utf8]{inputenc}
\usepackage{microtype}

\usepackage{graphicx}
\usepackage{subfigure}
\usepackage{amsmath}
\usepackage{amsfonts}
\usepackage{epstopdf}
\usepackage{color}
\usepackage{gensymb}
\usepackage[section]{placeins}
\usepackage[table]{xcolor}
\usepackage{array} 
\usepackage{longtable}
\usepackage{tabu} 
\usepackage{eurosym}

\usepackage{interval}

\usepackage{tabularx}
\usepackage{booktabs}   
\usepackage{multirow}
\newcommand{\ra}[1]{\renewcommand{\arraystretch}{#1}}
\newcommand{\specialcell}[2][c]{%
	\begin{tabular}[#1]{@{}l@{}}#2\end{tabular}}
\newcolumntype{R}[1]{>{\raggedleft\let\newline\\\arraybackslash\hspace{0pt}}m{#1}}
\usepackage{threeparttable}
\usepackage{colortbl}
\usepackage{longtable, booktabs, boldline}
\usepackage{cellspace}
\usepackage{textcomp}
\usepackage[export]{adjustbox}

\definecolor{Orange}{rgb}{0.9,0.5,0}
\definecolor{NavyBlue}{rgb}{0.1, 0.4, 0.8}
\definecolor{Magenta}{rgb}{0.8, 0.1, 0.6}
\definecolor{Cyan}{rgb}{0.2, 0.85, 0.85}

\usepackage{xr-hyper} 
\makeatletter
\newcommand*{\addFileDependency}[1]{
  \typeout{(#1)}
  \@addtofilelist{#1}
  \IfFileExists{#1}{}{\typeout{No file #1.}}
}
\makeatother

\usepackage{hyperref}
\usepackage{cleveref}

\usepackage{colortbl}
\definecolor{dispersionColor}{rgb}{0.517, 0.66, 0.714}
\definecolor{balanced1Color}{rgb}{0.870, 0.949, 0.765}
\definecolor{balanced2Color}{rgb}{0.99, 0.89, 0.529}
\definecolor{agglomerationColor}{rgb}{0.99, 0.48, 0.196}

\definecolor{bin1}{rgb}{0.74509804, 0.86666667, 0.8745098}
\definecolor{bin2}{rgb}{0.89803922, 0.67058824, 0.57254902}
\definecolor{bin3}{rgb}{0.84705882, 0.49803922, 0.4}
\definecolor{bin4}{rgb}{0.80784314, 0.34117647, 0.27058824}
\definecolor{bin5}{rgb}{0.70196078, 0.21960784, 0.2}
\definecolor{bin6}{rgb}{0.54117647, 0.18039216, 0.17254902}

\begin{document}

\title{The agglomeration and dispersion dichotomy of human settlements on Earth}
\author{Emanuele Strano$^*$}
\affiliation{MindEarth, 2502 Biel/Bienne, CH}

\author{Filippo Simini$^*$}
\affiliation{University of Bristol, 06010 Bristol, UK}

\author{Marco De Nadai$^*$}
\affiliation{Fondazione Bruno Kessler (FBK), 38123 Trento, Italy}
\affiliation{University of Trento, 38123 Trento, Italy}

\author{Thomas Esch}
\affiliation{German Aerospace Center (DLR), 82234 Wessling, Germany}

\author{Mattia Marconcini}
\affiliation{German Aerospace Center (DLR), 82234 Wessling, Germany}

\maketitle
\def\thefootnote{*}\footnotetext{These authors contributed equally to this work.}\def\thefootnote{\arabic{footnote}}

\textbf{Abstract}
Human settlements on Earth are scattered in a multitude of shapes, sizes and spatial arrangements. These patterns are often not random but a result of complex geographical, cultural, economic and historical processes that have profound human and ecological impacts. 
However, little is known about the global distribution of these patterns and the spatial forces that creates them.
This study analyses human settlements from high-resolution satellite imagery and provides a global classification of spatial patterns. We find two emerging classes, namely agglomeration and dispersion. In the former, settlements are fewer than expected based on the predictions of scaling theory, while an unexpectedly high number of settlements characterizes the latter. Our global classification of spatial patterns correlates with some urban outcomes, such as the amount of CO2 emitted for transportation, providing insights into the relationship between land use patterns and socio-economic and environmental indicators. To explain the observed spatial patterns, we also propose a model that combines two agglomeration forces and simulates human settlements' historical growth.
Our results show that our model accurately matches the observed global classification (F1: 0.73), helps to understand and estimate the growth of human settlements and, in turn, the distribution and physical dynamics of all human settlements on Earth, from small villages to cities.

\section*{Introduction}
The growth and expansion of cities on Earth influence all global social, economic and environmental systems~\cite{leung2020clustered, united2001,UN20045,UN2006,Birch2011}. 
Abundant evidence indicates that cities have significant impacts on the water and ecological systems, land-use competition, food production, biodiversity, climate change and human health~\cite{Moore2003,Zhou2004,kaufmann2007,Grimm756,Tilman2011}, and extensive debates highlight the trade-off between benefits and challenges for global urbanization~\cite{Daily1992,Johnson2001,Dye766,seto2011, d2017future, guneralp2017global}.
However, the real extent, distribution and explanation of human settlements (HSs) are not yet fully understood at the global scale, especially regarding the spatial arrangement and type of patterns for settlements of \textit{all sizes}, ranging from vast metropolitan areas to small and scattered rural settlements. 

Several factors have hampered a global analysis and description of HSs: on the one hand, quantitative analyses of HSs patterns often rely on traditional spatial metrics used in urban geography~\cite{herold2002use}, typically extracted from census data~\cite{barrington2015century, hamidi2014longitudinal, huang2007global, poelmans2009detecting}, and statistical analyses derived from complex systems such as fractals and scaling analysis~\cite{batty2008the-size}, which are observed only at large spatial scales such as continents and countries. 
On the other hand, most early studies relied on low- or medium-resolution satellite data that range from $0.5$ to $1$ km~\cite{Potere2007,gamba2009,grekousis2015overview, Angel2011,seto2011}, which are usually focused on \textit{urban} land cover and thus exclude from the analysis the vast majority of \textit{non-urban} settlements. 
Although high-resolution global HSs inventories have recently been proposed~\cite{Pesaresi2016, chen201730}, significant inaccuracies still exist~\cite{esch2017breaking}, probably due to the technical challenges of having a uniform and consistently cross-validated global dataset.

Here, we provide an unprecedented global estimation of the geography of HSs by quantitatively analyzing their location, distribution and spatial patterns through the scaling analysis based on the Zipf's law ~\cite{Zipf,rozenfeld2008laws,gabaix2004the-evolution,rybski2013distance-weighted}.
First, we provide a comprehensive global analysis of the location and density of all HSs by exploiting the World Settlement Footprint 2015 (WSF2015)~\cite{WSF_data_paper} dataset, an accurate 10 meters resolution inventory of human-occupied land. 
Second, we exploit scaling theory and analyze the deviations from the scale-free distribution of settlement sizes. We discover that in all continents two distinct types of HS patterns emerge: dispersed and agglomerated settlements.
These two patterns drive the high heterogeneity of HSs, help understanding urbanization in different areas of the world, and correlate with some urban outcomes such as the amount of CO2 emitted for transportation.
Finally, we build a minimal spatially explicit model that can reproduce all observed settlement patterns on Earth by interplaying two agglomeration forces.

\section*{Results}
We study HSs on Earth through the WSF2015 dataset~\cite{WSF_data_paper}, a novel $10$m resolution binary mask outlining the human-occupied land in the world. The dataset has been created by jointly exploiting multi-temporal radar (Sentinel-1) and optical (Landsat-8) satellite imagery, and it has been validated extensively~\cite{WSF_data_paper} through a collaboration between Google and DLR.

\begin{figure*}[ht!]
\centering
\includegraphics[width=\linewidth]{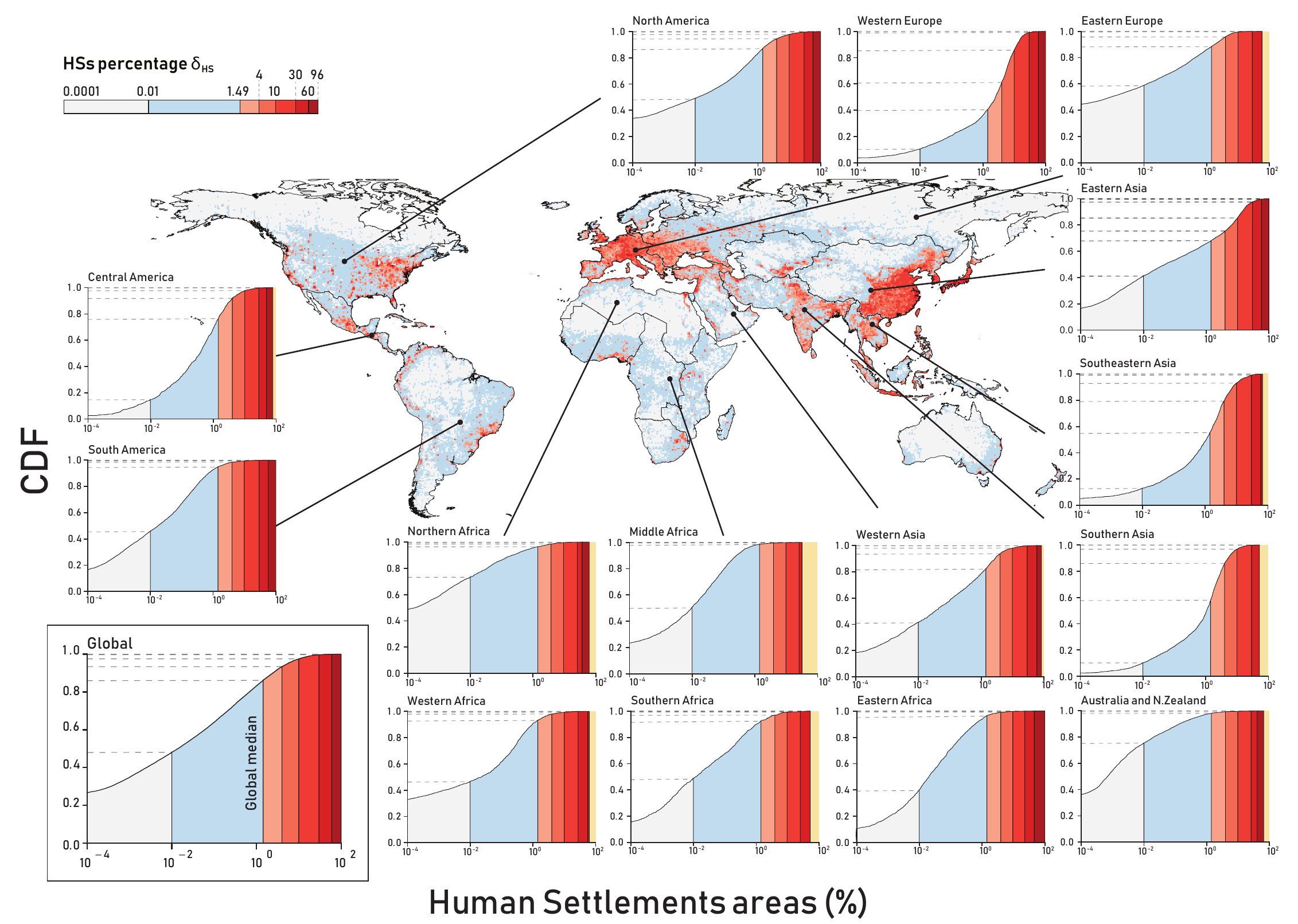}
\caption{A global overview of HSs density in 2015. For each tile, we compute the percentage of occupied HS area in each tile $\delta_{HS}$ and find that, on average, HSs cover 1.49\% of the tile area. Inset: the long-tail distribution of the cumulative frequency of the percentage of occupied HS area within each tile, $\delta_{HS}$. A small number of tiles contains the majority of the settlements. Map: we colour in cyan the tiles with $\delta_{HS}$ less than the global average density ($1.49\%$), while we colour with a red gradient the areas with a density between $1.49\%$ and $100\%$. High-density tiles are not evenly distributed in the world. Small plots: we show how the global long-tail emerge in all the UN-defined macro-areas.}
\label{fig2}
\end{figure*}

The WSF2015 classifies as human-occupied land a $10 \times 10$m cell that contains either a building or a building lot, where: i) a building is any structure having a roof supported by columns or walls and intended for the shelter, housing, or enclosure of any individual, animal, process, equipment, goods, or materials of any kind; and ii) a building lot is the area contained within an enclosure (e.g., wall, fence, hedge) surrounding a building or a group of buildings. 
Such an accurate inventory of human presence on Earth allows us to perform an unprecedented analysis of the real magnitude, geography and spatial structure of HSs at the global level.

From the WSF2015, we define an HS as a continuous
areas of human-occupied land formed by aggregating neighbouring pixels whenever one touches the other along its edges (see the Methods Section for details). Thus, an HS might be as small as a single building or big as an entire city.

We estimate that the total number of HSs is approximately $32$ million and the corresponding area amounts to 1,302,187 km$^2$ (i.e., about $1.04\%$ of the global land surface area estimated in 131,331,424 km$^2$ excluding the Arctic and Antarctic regions). 
However, not all dry-land surfaces can be settled. Thus, from satellite imagery, we exclude areas with complex topography that are not suitable for hosting HSs (e.g. areas with extremely elevated steepness) and internal freshwater surfaces through a \textit{relief mask} and a \textit{freshwater mask}, respectively (see the  Methods and Supporting Information (SI) sections for details).
The area of habitable land amounts to 106,445,525 km$^2$; out of this, we estimate that HSs cover $1.22\%$ of the entire world.


However, settlements on Earth are not evenly distributed across regions, and they are very heterogeneous in size and shape.
To study such variations, we subdivided the Earth's surface into 29,181 tiles of $0.5^{\circ} \times 0.5^{\circ}$ (approximately $55 \times 55$ km$^2$ at the equator). We measured the percentage of occupied HS area $\delta_{HS}$, or density, in each tile as the ratio between the tile's HS area and its total surface area minus the exclusion mask (defined as the combination of the relief and freshwater areas) and find that, on average, HSs occupy 1.49\% of the tile's area.
Figure~\ref{fig2} shows the spatial distribution and cumulative frequency of $\delta_{HS}$ at the global scale, and for the 16 macro-areas defined by the United Nations~\cite{united1982standard}.
In the bottom-left inset of Figure~\ref{fig2} we plotted cumulative frequency at the global scale by fixing on the $x$-axis seven HS percentage thresholds. We find that the density of HSs areas on Earth has a long-tail distribution, which means that a small number of tiles contains the majority of the settlements while there are many tiles with few HSs.

\subsection*{Density-independent classes of human settlements' patterns}

\begin{figure*}[ht]
\centering
\includegraphics[width=0.9\linewidth]{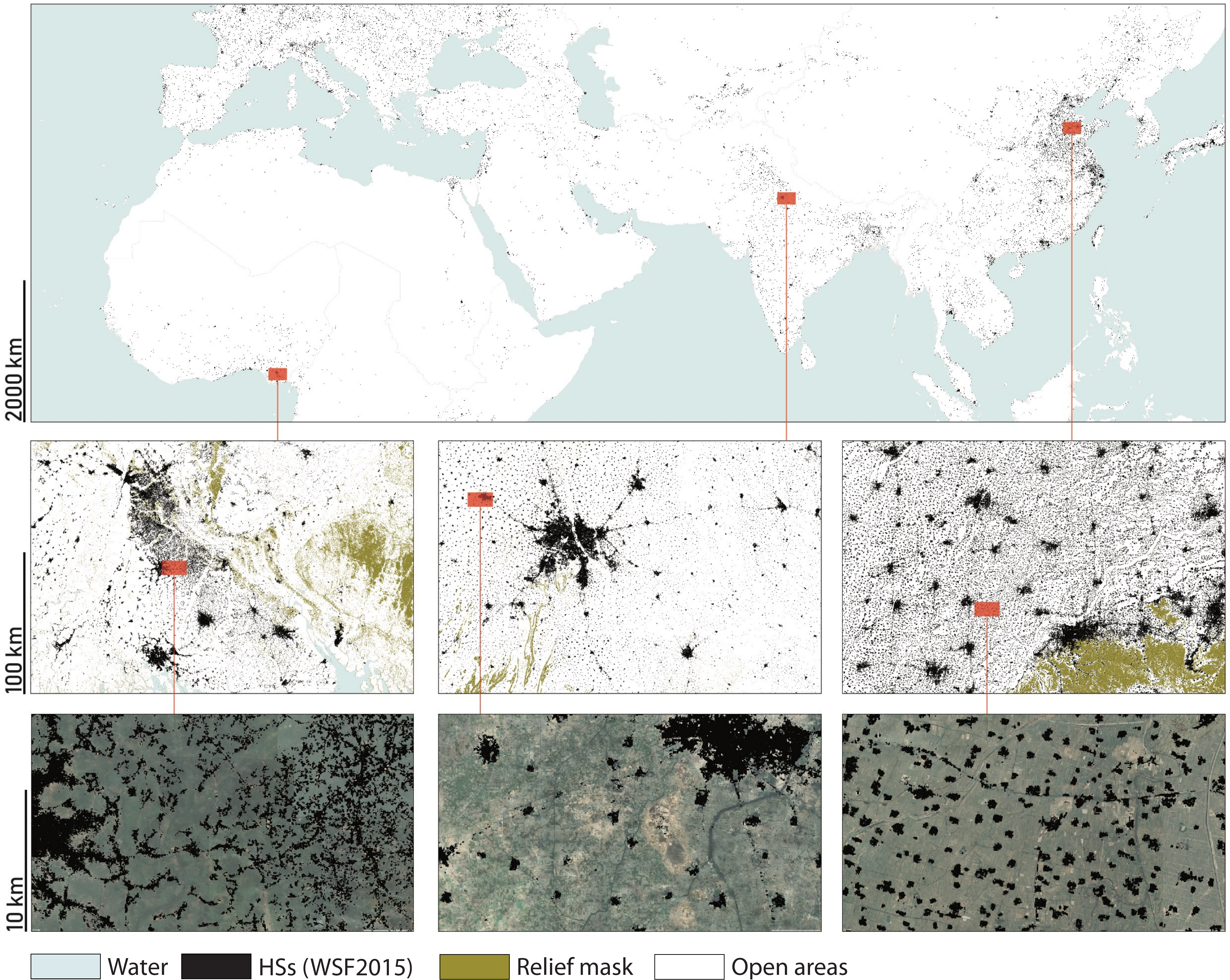}
\caption{Three examples of the high heterogeneity of spatial patterns referring to the Igboland (Nigeria), New Delhi (India) and Jinan (China) regions. The HSs computed from the WSF2015 are shown in black, the relief mask in light brown, the water mask in blue, and the remaining open areas (including natural and cropland areas) in light green. In the bottom row, the background shows HR optical satellite imagery.} 
\label{fig1}
\end{figure*}

The spatial distribution of density alone does not explain the complexity of HSs patterns on Earth (see Figure~\ref{fig1}), which are very heterogeneous in shape and dimension. Such variety of patterns may arise from the very well-known spatial interpenetration of rural and urban settlements~\cite{gottmann1957megalopolis}, which results in a complexity of shapes and sizes that no longer fit those classes. This phenomenon has been qualitatively observed in classical urban geography narratives through the notions of \emph{megalopolises}~\cite{gottmann1957megalopolis}, \emph{urban sprawl}~\cite{indovina1990citta} and \emph{horizontal metropolises}~\cite{vigano2017rethinking}.
However, this gradual symbiosis of different urbanization forces has never been quantitatively defined and tested.
We here propose a quantitative classification of settlement patterns based on scaling analysis~\cite{christensen2005complexity,stanley1999scaling}. 

In the context of urbanization and HSs patterns analyses, some invariant spatial proprieties of HSs~\cite{batty2008the-size} and transportation networks~\cite{Str_NR} have been found to follow scale-free relationships.
The strongest empirical evidence of a power-law relationship in urban science is the scale-free distribution of settlement sizes: the probability of observing a settlement with an area larger than $A$ follows a power law, $P(A) \sim A^{-\alpha}$, also called Zipf's law~\cite{Zipf,rozenfeld2008laws,gabaix2004the-evolution,rybski2013distance-weighted}.
Accordingly, the areas of the HSs in the tile and those in its corresponding UN-defined macro-area $m$ are expected to be sampled from the same empirical distribution, $P_{m}(A)$, which is well approximated by the Zipf's law as expected (see SI, Figure S1).
Based on this assumption, for each HS $i$ in a $0.5^{\circ} \times 0.5^{\circ}$ tile, we measure its area $A_{i}$ and the total HS area of a tile $A^{tot}_{HS} = \sum_{i=1}^N A_{i}$, where $N$ is the number of HSs in the tile. 
Then, for each tile in macro-area $m$ with a total settlement area $A^{tot}_{HS}$, we estimate $P_m(N|A^{tot}_{HS})$ following~\cite{simini2019testing}.
To do so, we randomly sample the HS areas from $P_{m}(A)$ until the sum of the sampled areas is equal to $A^{tot}_{HS}$ and find the number of HSs $N$ we sampled.
Then, we estimate the distribution $P(N | A^{tot}_{HS})$ by repeating the process 1000 times (see Methods).
Note that the expected number of HSs increases with the total target area $A^{tot}_{HS}$.
If the observed values of the number $N$ of HSs is distributed according to the theoretical distribution $P_m(N | A^{tot}_{HS})$, then the corresponding quantiles $Q(N) = F(N | A^{tot}_{HS})$ should be distributed uniformly between 0 and 1, where $F$ is the cumulative distribution of $N$.

\begin{figure}[ht]
\centering
\includegraphics[width=0.6\linewidth]{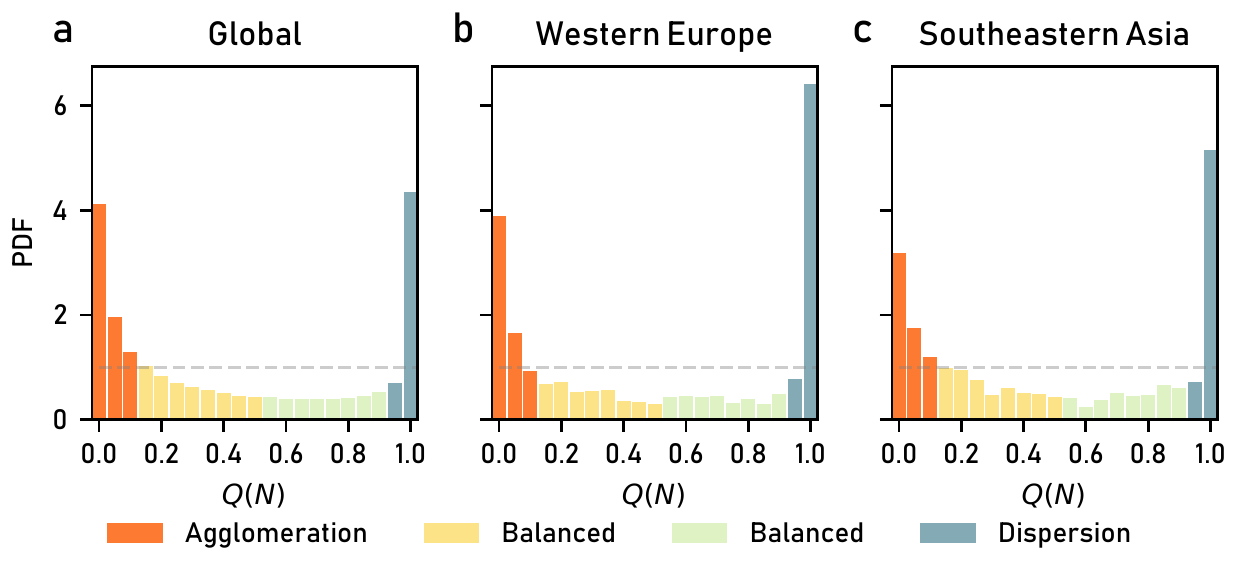}
\caption{
The distribution of the deviations from the Zipf's law. Deviations with $Q(N)\simeq~0$ correspond to tiles with a smaller number of HSs than expected, while when $Q(N)\simeq~1$ tiles have a higher number of HSs than expected by the theoretical model. We observe two peaks in all the macroareas. a) the global distribution; b-c) two examples of distributions within a macroarea. The colors indicate our proposed classification.
}
\label{fig:bimodality}
\end{figure}

However, we find that the empirical quantiles are not uniformly distributed between 0 and 1. Instead, we observe a bimodal distribution with two distinct peaks located around $Q(N)=0$ and $Q(N)=1$ (see \Cref{fig:bimodality}). Similar results are observed in most macro-areas (see SI Figure S2).
Thus, based on the theoretical quantiles $Q(N)$, we define two extreme classes of settlement patterns: a {\it dispersion class} ($0.9 \leq Q(N) < 1 $, 10th decile), corresponding to tiles with a large number of HSs according to the theoretical expectations; and an {\it agglomeration class} ($0.0 \leq Q(N) < 0.1 $, 1st decile), corresponding to tiles with a small number of HSs according to the theoretical expectations.
In between these two extreme classes, we define the {\it balanced class} ($0.1 \leq Q(N) < 0.9 $, 2nd-9th deciles), divided into two sub-groups ($0.1 \leq Q(N) < 0.5 $ and $0.5 \leq Q(N) < 0.9$) to better understand the patterns of the tiles.

\begin{figure*}[ht]
\centering
\includegraphics[width=\linewidth]{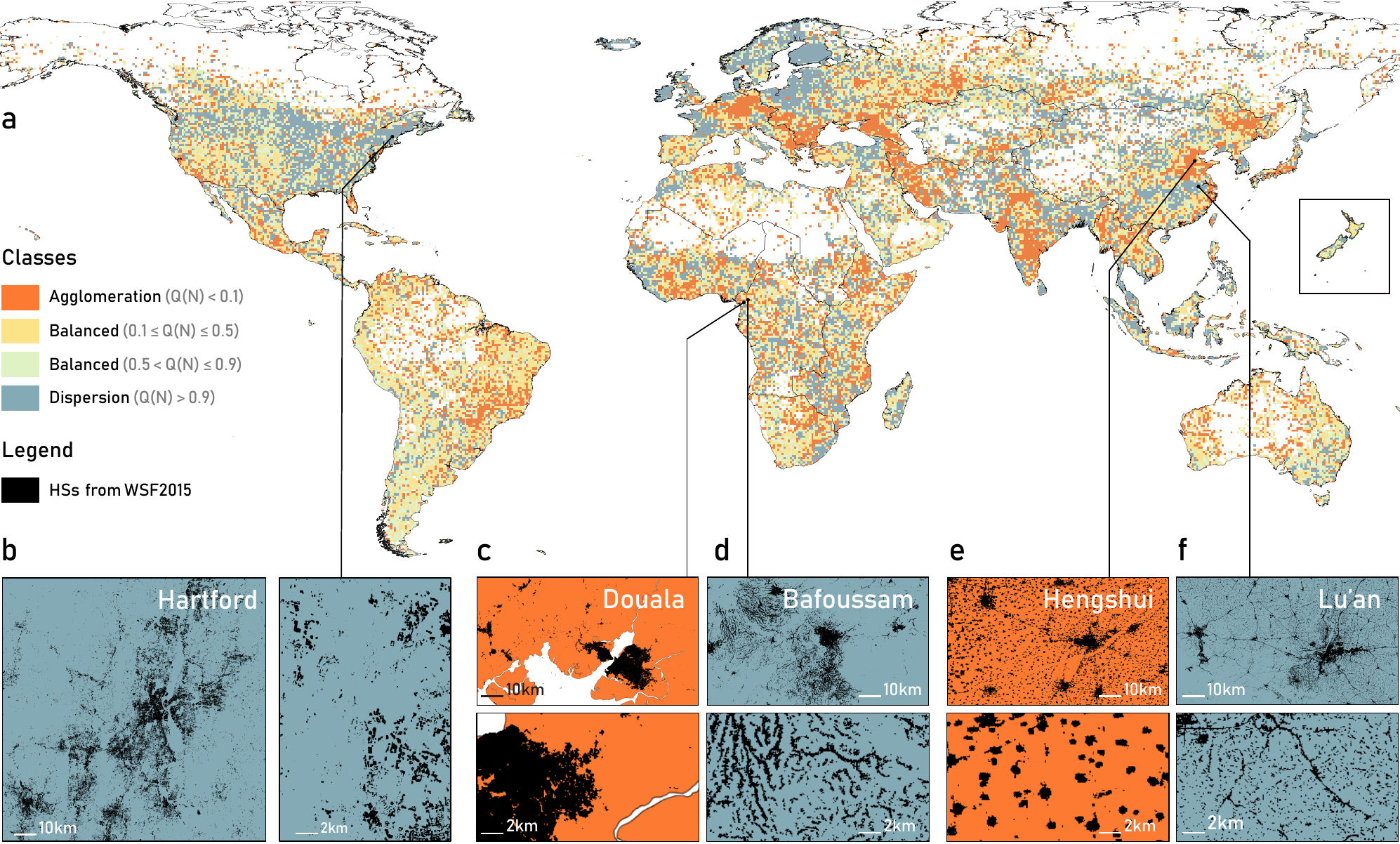}
\caption{Global classification of the HS patterns by their deviation from the scaling analysis predictions. The colour range is consistent across all panels. The blue insets (b, d, and f) show the \textit{dispersion} class, which represents locations with more HSs than expected from the scaling theory. b) Two different zoom levels of Hartford's metropolitan area (USA), where urban dispersion is due to extensive patterns of single housing and car-centric transportation. d) Urban-rural agglomeration in the area around Bafoussam (Cameroon), where settlements are mostly composed of informal single housing/agricultural units. f) Area around Lu'an (China), where there is an abundance of low-density sparse and small settlements.
The orange insets (c and e) show the \textit{agglomeration} class, in which we find fewer settlements than expected. c) The city of Douala (Cameroon), where urbanization occurs tightly around the existing urban core. e) The area around the city of Hengshui (China), where the many small settlements are most likely due to the agricultural land preservation strategies that have been extensively developed with a focus on compactness.}
\label{fig3}
\end{figure*}

Figure~\ref{fig3}a shows the spatial distribution of the classified tiles at a global scale.
We observe that the tiles classes are not spatially distributed at random, but they tend to form spatially compact clusters. For example, the blue cluster in the dispersion class in southern China (in Figure~\ref{fig3}c) and the orange cluster in the \emph{agglomeration} class in northern China (in Figure~\ref{fig3}e) are of considerable size and consist of multiple tiles. 
The fact that the classes of settlement patterns are not randomly distributed in space shows that the proposed classification scheme captures patterns characterizing large geographical regions and possibly large urban corridors. 
In North America, more precisely in the United States (US), we notice a large number of tiles in the dispersion class (blue tiles in \Cref{fig3}a and  \Cref{fig3}b), whereas the rest of the tiles in the US are mostly within the \emph{balanced} class (light yellow and green tiles), except for a few large urban agglomerations in the \emph{agglomeration} class (orange tiles).
This picture is in agreement with recent measurements of urban sprawl in US metropolitan areas and counties, which was evaluated using factors such as development density, land use mix, activity centring, and street accessibility~\cite{hamidi2014longitudinal}.

The proposed classification can highlight similarities and differences of HSs patterns observed on Earth.
We find, for example, 
that highly compact cities, such as Douala, Cameroon (see \Cref{fig3}c), belong to the same class of highly saturated areas like the city of Hengshui, China (see \Cref{fig3}e). These areas may appear different at first glance; however, they are intrinsically similar, as in both cases, the settlements are compact, regardless of their spatial distribution.
This classification is corroborated by a qualitative understanding of these two areas: Douala probably attracted all new settlers around the urban core as it is a port town and the wealthiest and most industrialized town in Cameroon, whereas, near Hengshui, the over-abundance of compactly developed settlements is due to avoidance of excessive erosion of productive agricultural land. By contrast, the Lu'an region (see \Cref{fig3}f), which is also an agricultural area, belongs to the dispersion class probably because it has not been regulated by agricultural land erosion protection policies and thus presents a highly dispersed pattern of settlements. The same highly sprawled pattern appears in several mega-settlement agglomerations in sub-Saharan Africa, where large sub-urban areas are dominated by single-plot housing as in the area of Bafoussam, Cameroon (Figure~\ref{fig3}d).

We also observe that the observed bimodality in the deviations from scaling theory predictions cannot be explained by and is not a simple by-product of a different distribution of HS sizes for those tiles.
The distributions of HS sizes for tiles in the dispersion class are indeed not systematically different from the distributions of the tiles in the \emph{balanced} class (see SI, Figure S3).
The excess of tiles in the \emph{dispersion} and agglomeration classes is observed across all values of the percentage of HS areas, $A^{tot}_{HS}$, indicating that an over-abundance of HSs is not specific to lowly or highly urbanized regions and is thus independent of urban density (see SI, Figure S4).

We also examine the connection of some urban indicators with the deviations from the theoretical distribution. Particularly, we consider the on-road $CO_2$ emissions, road length, use of the car and public transportation for commuting in the USA (see Materials and Methods).
We fit each urban indicators controlling by $Q(N)$ and the total HS area of each tile through a multivariate Ordinary Least Squares regression. 
We find that higher dispersion (i.e. higher $Q(N)$) seems to indicate higher levels of on-road $CO_2$ emissions and on-road $CO_2$ emissions per capita. Similar trends also exist for road length per capita and higher private cars usage for commuting (see SI Table S5).
Moreover, we find that the more aggregated the HSs are, the higher is the usage of public transportation (see SI Table S5). 
We note that the goodness of fit of all urban indicators significantly decrease (on average by 41\%) when we consider the HSs total area as the only independent variable in the regression. Similar results are obtained with non-linear regressions (see SI Table S5).
Not only these correlations confirm previous results on the connection between urban sprawl and pollution~\cite{ribeiro2019effects, gately2015cities, stone2008urban} but it allows to draw conclusions without relying on the large amount of hand-crafted spatial metrics proposed in literature~\cite{huang2007global, hamidi2014longitudinal, barrington2015century, poelmans2009detecting}.

\subsection*{A spatial model for human settlements}
The deviations from scaling theory predictions show a high heterogeneity of HSs patterns, resulting from numerous historical dynamics.
In the absence of global and precise historical HSs data, we shed light on how such a variety might be achieved through controlled spatial simulations. 

We hypothesize that HSs evolution cannot be attributed to agglomerating forces alone but rather to more complicated systems of spatial forces. To test this hypothesis, we here propose an extension of random percolation models~\cite{makse1998modeling} to simulate and reproduce such a system of forces and explain the macro-dynamics in action during settlement evolution. 
Our proposed model works in a two-dimensional lattice $w$ of size $L \times L$, where $L=1000$, whose sites (cells) can be either occupied ($w_{i,j} = 1$, human settlement (HS)) or empty ($w_{i,j} = 0$, undeveloped).
Without loss of generality, the initial configuration has only the central cell occupied ($w_{L/2, L/2} = 1$), and all other cells are empty ($w_{i, j} = 0\ \forall i,j \neq L/2$). Then, the model iteratively simulates the growth of settlements; at each step, the probability that each empty cell is occupied is:
\begin{equation*}
        q_{i,j} = 
        C \hat{q}_{i,j}
        = C \frac{\sum_{k}^L\sum_{z}^L w_{k,z} d_{ij,kz}^{-\gamma}}{\sum_{k}^L \sum_{z}^L d_{ij,kz}^{-\gamma}}
\end{equation*}
where $C = 1 / \max_{i,j}(\hat{q}_{i,j})$ is a normalization constant and $d_{ij,kz}$ is the Euclidean distance between site $w_{i,j}$ and site $w_{k,z}$.

As in its traditional form~\cite{rybski2013distance-weighted}, the parameter $\gamma$ regulates the strength of attraction of an HS cell on a new cell; $\gamma = 0$ implies a dispersed and randomly located new occupied cell, while a larger $\gamma$ attracts new cells close to old cells, thus producing mono-centric and agglomerated patterns (see SI, Figure S7). 
To simulate the different forces in action, the simulation is split in two steps that are controlled by the parameter $s$. When the fraction of the occupied cells in the simulation is less than a given percentage $s$ (i.e., $\sum_{i,j} w_{i,j} / L^2 \leq s$) $\gamma = \gamma_1$, while $\gamma = \gamma_2$ when the fraction of occupied cells is greater than $s$:
\begin{equation*}
    \gamma=
    \begin{cases}
      \gamma_1, & \text{if}\ \sum_{i,j} w_{i,j} / L^2 \leq s\\
      \gamma_2, & \text{otherwise}
    \end{cases}
\end{equation*}
\noindent 
$\gamma_1$ characterises settlements' expansion during the initial stages of the simulation while $\gamma_2$ characterises settlement expansion for the rest of the simulation (see SI, Figure S8 for a visual explanation of the patterns generated).
The model, which we call a multi-parameter model, has three parameters: $\gamma_1$, $\gamma_2$ and $s$. When $\gamma_1 = \gamma_2$, it becomes equivalent to the single-parameter model~\cite{rybski2013distance-weighted}.

\begin{figure*}[ht]
\centering
\includegraphics[width=\linewidth]{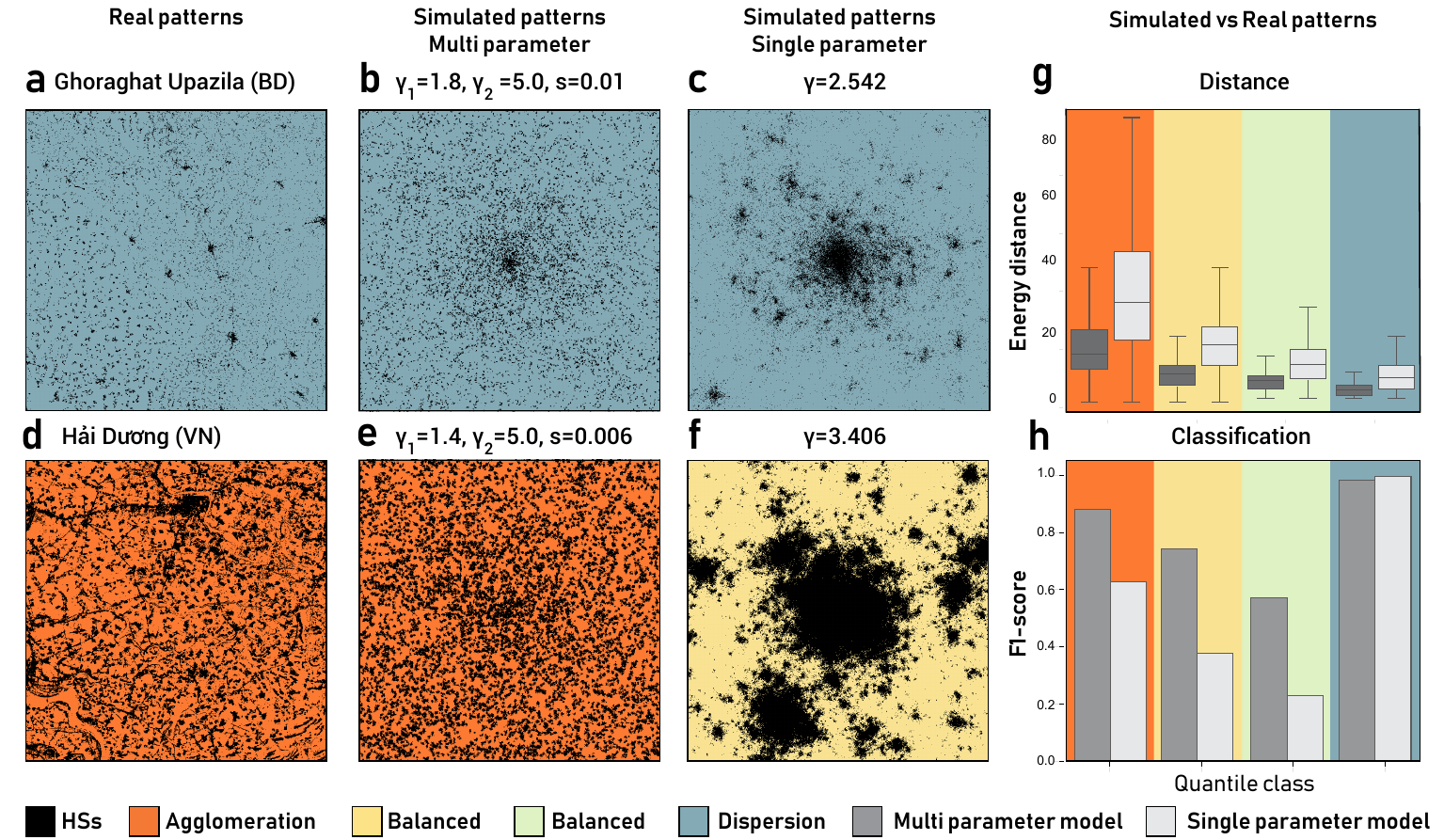}
\caption{
Qualitative and quantitative results of the proposed multi parameter model. a) Area near Ghoraghat Upazila, Bangladesh, having a settlement pattern in the \emph{dispersion} class; b) the most similar simulation obtained with our model; c) although the most similar simulation obtained with the single-parameter model lies in the same class as the real tile, it has a very different pattern. 
d) Area near Hai Duong, Vietnam, having a settlement pattern in the \emph{agglomeration} class; e) the most similar simulation obtained with our model; f) the most similar simulation obtained with the single-parameter model falls into the wrong class. g) This box plot shows the energy distance between the real and simulated tiles computed for each class. It shows that our model always generates settlement patterns that are consistently better than those generated by the single-parameter model. h) The F1-score obtained from the urbanization class of the real tile and the class of its most similar simulation for both our model and the single-parameter model. Our model outperforms the single-parameter model and generates settlement patterns compatible with the classes observed in the real world.
}
\label{fig4}
\end{figure*}

We follow a simulation approach in which we find the parameters that best represent the spatial process that might have generated the patterns of the real tiles. First, we generate approximately 1,000,000 simulations using a broad range of parameter values (see SI, Table I) and simulate patterns until the lattice reaches 60\% of occupied cells. 
For each real tile $i$, we find the most similar simulation by comparing the cumulative distributions of HS sizes and selecting the simulated tile with the smallest Wasserstein distance $D_{E}(i)$~\cite{vaserstein1969markov} between the distribution of HS sizes of real and simulated tiles (see Methods). 
Finally, for each simulated tile, we also find its class of settlement patterns (i.e. \emph{agglomeration}, \emph{balanced}, \emph{dispersion}) by the quantile procedure mentioned before.
\Cref{fig4}a shows a randomly chosen tile in Ghoraghat Upazila, Bangladesh, while \Cref{fig4}b shows its most similar simulation with parameters $\gamma_1 = 1.8$, $\gamma_2 = 5.0$ and $s=0.01$. This simulation describes the dispersal phase of the real tile well in both its HS pattern and the class of settlement patterns. The same cannot be said for the most similar simulation from the single-parameter model, as it fails to describe both the sprawled pattern (see \Cref{fig4}c).
Similarly, we see from \Cref{fig4}d and \Cref{fig4}e that the randomly chosen tile of Hai Duong, Vietnam, is very well described by our model with parameters $\gamma_1 = 1.4$, $\gamma_2 = 5.0$ and $s=0.006$, while the best simulation of the single-parameter model fails to simulate this large number of settlements in the \emph{agglomeration} class and its class(\Cref{fig4}f). More qualitative examples can be found in SI Figure S9-S16.

To perform a quantitative evaluation of the performance of the multi-parameter model, we assess its ability to generate realistic distributions of HS sizes and urbanization classes.
First, we compare the distributions of the Wasserstein distances $D_E$ across all the simulated tiles from the multi-parameter model and the single-parameter model (see \Cref{fig4}g).
The two-sided Kolmogorov-Smirnov (KS) test~\cite{kendall1938new} shows that the multi-parameter model has a significantly smaller distance for all urbanization classes (see SI, Table S1), with 45.85\% smaller median distances, on average.
This result is robust against different distance metrics (see SI for additional details).
Second, we compare the urbanization class of each real tile with the one of its most similar simulation.
We use the F1-score to quantify the agreement between the urbanization classes of the real and simulated tiles. We find that the multi-parameter model achieves 50.68\% higher performance than that of the single-parameter model (see SI, Table S2). \Cref{fig4}h shows that this increase in performance is evident for the \emph{balanced} and \emph{agglomeration} classes.
We found that the single-parameter model overestimates the number of tiles in the \emph{dispersion} class, while the multi-parameter model better captures the whole distribution of urbanization classes (see also SI Figure S6).
Moreover, we found out that the multi-parameter correctly simulates also the agglomeration-dispersion dichotomy we found in real data (see SI Figure S17).

\section*{Conclusion}
Due to global population growth, HSs are expected to increase accordingly. For this reason, the scientific understanding of the spatial patterns of HSs is of paramount importance for planning, managing, and eventually forecasting HSs and their consequences.

In this paper, we provide an unprecedented description of the geography and the spatial structure of all HSs on Earth. 
First, we exploit the state of the global art dataset of human-occupied land to reliably measure the location and distribution of all the land occupied by HSs. 
We find that the density of HSs areas on Earth has a long-tail distribution: very few zones on Earth are occupied by highly dense areas, while the vast majority of Earth is occupied by low-density scattered settlements composed of less than $2\%$ of HSs area.
These low-density and scattered patterns are not only the result of the expansion of metropolitan areas; they also depend on a different process that goes beyond the arbitrary rural-urban dichotomy. 
Cities are undoubtedly important to study for their socio-economic importance and \emph{agglomeration} effects~\cite{bettencourt2007growth, gomez2016explaining}. However, the long-tail distribution we find shows that the over-abundance of low-density areas occupy approximately 50\% of the global surface, and may deserve more attention from the scientific community.

Second, we show that settlement density alone does not explain the great variability of HS patterns on a global scale. 
Thus, we exploit scaling analysis to study the number of settlements expected to be found in a region with a given HS area. 
From the deviations of the scaling analysis predictions, two distinct classes of settlement patterns emerge, which we named \emph{dispersion} and \emph{aggregation}. The former contains regions with the highest number of settlements with respect to their HS area, according to the deviations from scaling analysis; conversely, the \emph{agglomeration} class contains regions with the smallest number of settlements with respect to their HS area, according to the predictions of the scaling analysis. 
We name the patterns between \emph{aggregation} and \emph{dispersion} as \emph{balanced}.
The deviations from scaling analysis predictions well describe some urban transportation metrics, and seem to indicate that dispersed areas produce more on-road $CO_2$ emissions, require more streets per capita and rely more on cars for commuting.
Thus, our global classification allows to understand and group the different patterns of HSs on Earth and might help better planning future policies for sustainable settlements' growth.

Regarding the deviations from scaling analysis, one can speculate that Zipf's law is not fully capable of describing urban patterns. We instead argue that scaling analysis is a valuable framework. We showed how deviations from Zipf's law could be used to produce a quantitative classification of HS patterns, which provides additional insights to policy-makers and goes beyond the traditional rural-urban dichotomy.

Finally, we propose a spatially explicit model to shed light on the process that might result in the observed HSs patterns, in the absence of time-varying data at a global scale. 
The tiles we simulate match well with the HSs patterns and classifications, both locally (\Cref{fig3}) and globally (see SI Figure S6).
The model is validated on multiple distance metrics and alternative baselines.
Our findings show that the spatial dynamical process that regulates attractive and dispersal forces while settlements grow may be subject to random processes and that their combinations are undoubtedly subject to local and specific conditions. As such, local and regional conditions must be taken into account when studying and modelling urban phenomena. 

A global and precise analysis of HSs does not come without limitations. It is worth noting that, due to limitations specific to the data used, it was not feasible to consistently and systematically detect globally tiny structures (e.g., huts, shacks, tents) due to their reduced scale, temporal nature (e.g., nomad or refugee camps), building material (e.g., cob, mud bricks, sod, straw, fabric), or the presence of dense vegetation preventing their identification. 
Moreover, we acknowledge that our simulations through the spatially explicit model find only a possible explanation for HSs' observed patterns and classification on Earth. We stress the need for a global, precise and reliable time-varying dataset of HSs to better understand the spatial processes underlying HSs' growth.

In our view, the analysis and model we propose represent a fundamental tool to provide insights about the structure and the evolution of HSs on Earth and, in turn, of their impact on humans and the environment. 
In the future, the observation of the Earth surface will experience tremendous improvement, providing more data that are more accurate and denser in time. We hope that our framework will pave the way to new research to understand the extent of HSs and manage better their impact on the environment and life on Earth.

\section*{Methods}
In this section we first describe how we delineate the HSs from satellite data, then we explain the relief mask and the segmentation process. Finally, we describe how we use the data to perform scaling analysis, the simulations and the comparisons with the real data.

\subsection*{Global HSs delineation from satellite imagery}
We exploit the World Settlement Footprint 2015 dataset~\cite{WSF_data_paper} to reliably and accurately outline HSs globally. This dataset is composed of multiple binary raster files obtained from 2014-2015 multi-temporal Sentinel-1 radar and Landsat-8 optical imagery (of which approximately 107,000 and 217,000 scenes were processed, respectively).
The dataset has an average resolution of 10 meters, and it has been tested in close collaboration with Google for a collection of 50 globally distributed test sites (tiles of $1 \times 1$ lat/long degree), including 900,000 reference samples. 

Physical environmental conditions play a significant role in HSs development; among these, terrain steepness is one of the most critical. Accordingly, to exclude from our analysis relief areas that are unfavourable for settlement, we generated – based on extensive empirical analysis - a binary mask using the Shuttle Radar Topography Mission (SRTM) Digital Elevation Model (DEM) available between -60° and +60° and the Advanced Spaceborne Thermal Emission and Reflection Radiometer (ASTER) DEM elsewhere. Specifically, the mask is positive where the shaded relief (depicting how the three-dimensional surface would be illuminated from a point light source) is greater than 212, or the roughness (defined as the largest inter-cell difference of a central pixel and its surrounding 8 cells) is greater than 15.


\subsection*{Global vectorial HSs}
Global urbanization is measured by taking into account HSs, water, and impervious areas. To facilitate the analysis at the global scale, the globe has been divided into a grid of $0.5 \times 0.5$ degrees in the 
{European Petroleum Survey Group (EPSG) 4326} projection. Using a global water mask, we select only the cells that intersect the emerged lands, which results in 63,507 cells available for the analysis.
First, we transform the raster databases into polygons at each cell through the GDAL 2.2.2~
and PostGIS software packages.
Next, we create a hierarchy of encapsulated grids where, at each level, a cell is composed of the four cells from the lower level (e.g., each cell of the $1 \times 1$ degree grid comprises four cells belonging to the $0.5 \times 0.5$ degree grid). At each level, the polygons are then merged on the boundaries of the lower level's cells. The result is a series of layers where urbanization can be analyzed and processed worldwide at multiple scales.

The HSs, water and impervious areas are calculated in kilometres through the Universal Transverse Mercator (UTM) projections.

\subsection*{Scaling analysis}
To numerically estimate the theoretical confidence intervals for the number of settlements $N$ predicted by scaling theory, we proceed as follows. 
We evaluate the theoretical conditional distribution of the number of settlements in a tile of total HS area $A^{tot}_{HS}$, $P_m(N|A^{tot}_{HS})$, by sampling with replacement from the list of settlement areas belonging to the tile's macro area $m$ until the total HS area (i.e., the sum of the sampled areas) is equal to the target value $A^{tot}_{HS}$. 
The number of samples $N$ needed to reach $A^{tot}_{HS}$ can be considered to be a number sampled from $P_m(N|A^{tot}_{HS})$. 
By repeating the sampling process 1000 times, we can evaluate the 1st and 9th deciles, corresponding to the boundaries of the \textit{agglomeration} and \textit{dispersion} classes, respectively. 

\subsection*{Correlations with emission and road data}
We focused on the USA to correlate the quantile from the theoretical distribution with the roads length and on-road CO2 emissions. 
The former data is extracted from OpenStreetMap, which is a collaborative project to map the world. The latter comes from DARTE Annual On-road CO2 Emissions dataset~\cite{gately2019darte} that is a 1-km resolution inventory estimated by the Federal Highway Administration's (FHWA's) Highway Performance Monitoring System (HPMS). We expect the OpenStreetMap data to have high quality in the USA. 
To compute the per-capita metrics we used the population estimates from Worldpop~\cite{worldpop2018global}.

Additionally, we also extracted the modal share of commuting trips for all the US tracts from the American Community Survey (ACS) (\url{https://www.census.gov/programs-surveys/acs}), which provides the percentage of commuting trips in terms of transportation mode.

\subsection*{Evaluation of the multi-parameter model}
Estimating the urbanization process would require temporal data, which are not easy to obtain. Moreover, a model fit on temporal data, where each pixel value is related to all the other pixels through a distance matrix, would be very computationally expensive.
Indeed, each tile contains $n = 5567 \times 5567$ cells, and a full distance matrix would require $\mathcal{O}(n^2)$ memory.
For this reason, we evaluate our model through simulations.

First, we simulate $1000 \times 1000$ tiles with an exhaustive grid search created from the Cartesian product of the ``reasonable'' values chosen for $\gamma_1, \gamma_2, s$ (see SI, Table S3). The set of all the simulation tiles is denoted by $\mathcal{S}$. 
Next, we compare the resulting simulations with the global (\emph{real}) tiles.
For each tile with an urbanization percentage $U_r \ge 1\%$, we find the simulated tile that is most similar to it by finding all the simulated tiles with an urbanization percentage $U_s \in \interval{U_r - 0.5\%}{U_r + 0.5\%}$. We compare the tiles via the Wasserstein distance $D(X_i, X_j)$, which is also known as Earth mover's distance, between the distributions $X_i$ and $X_j$ of HSs areas in the real and simulated tiles, respectively.
We denote the distance of a tile $i$ to its most similar simulated tile by $D_E(i) = \min_{j \in \mathcal{S}} D(X_i, X_j)$. We also tested other distance measures but did not find significant differences (see SI Figure S5).
As the size of the simulations is $1000 \times 1000$ pixels, we resize the real tiles to the same dimension with a nearest-neighbour approach before applying the distance function. This resizing makes the tiles comparable.

For each pair $(r, s)$, where $r$ is the real $1000 \times 1000$ pixel tile and $s$ is the simulated $1000 \times 1000$ pixel tile, we compute the quantile class of $r$ and $s$ and frame it as a classification problem. We compute the F1-score between the ground truth (the classes of the real tiles) and the predicted classes (the classes of the simulated tiles). The F1-score for all the classes is weighted to account for the unbalanced number of tiles in each class.

\section*{Data and code availability}
This research is based upon data openly available in the Internet. We release the code and the instructions to download all the source and intermediate data to repeat all the analysis and replicate the figures at \url{https://github.com/denadai2/precise-mapping-human-settlements}.

\section*{Author contributions}
All authors conceived the study. M.M. produced the World Settlements Footprint. F.S., M.D.N. and E.S. conducted the analyses and analyzed the results. F.S., M.D.N. and E.S. wrote the paper. All authors reviewed the manuscript.

\section*{Competing interests}
All authors declare no competing interests. 

\section*{Materials \& Correspondence}
Correspondence and material requests can be addressed to: \url{denadai@fbk.eu}

\section*{Acknowledment}
This work was partially supported by the Microsoft Azure Research Award. F.S. is supported by the  EPSRC First Grant EP/P012906/1.
Parts of the spatial metrics assessment and applications for this study were funded by the European Space Agency (ESA) under the Urban Thematic Exploitation Platform project (TEP Urban, ESRIN/Contract No. 4000113707/15/I-NB). E.S. thanks Enrico Bertuzzo and Marta Gonzalez. M.D.N. and E.S. thanks Nicu Sebe and Bruno Lepri.

\bibliographystyle{naturemag}
\bibliography{biblio}

\clearpage
\appendix

\section*{Additional figures}
\begin{figure}[ht]
\centering
\includegraphics[width=0.9\textwidth]{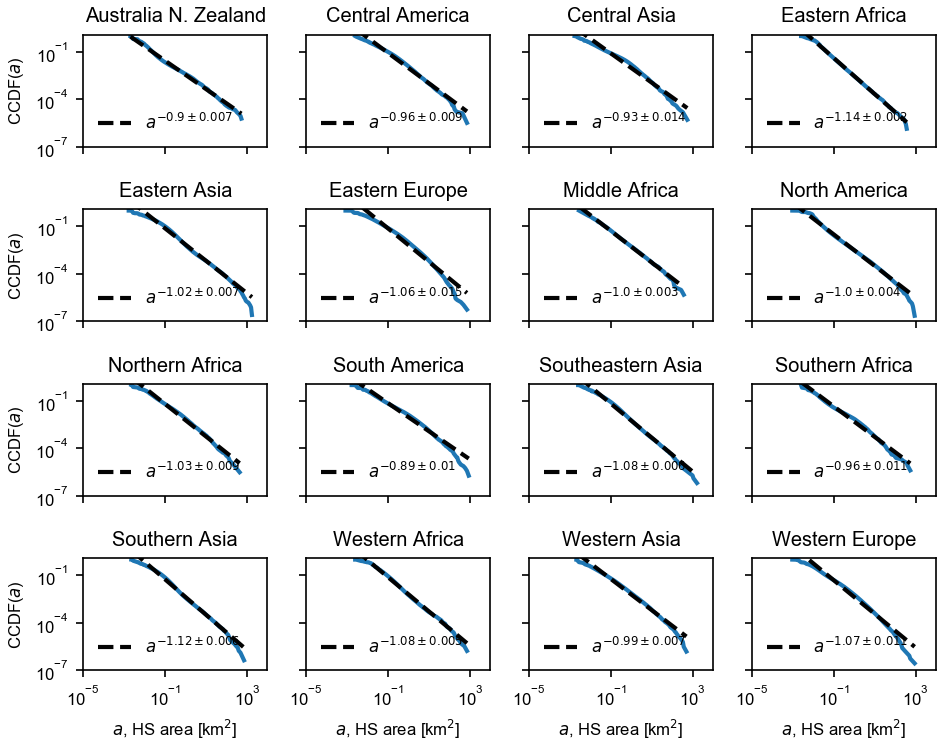}
\caption{Counter cumulative distribution function (CCDF) of the HS areas for the 16 macro regions considered.}
\label{fig:appendix:ccdf_areas}
\end{figure}

\begin{figure}[ht]
\centering
\includegraphics[width=0.7\textwidth]{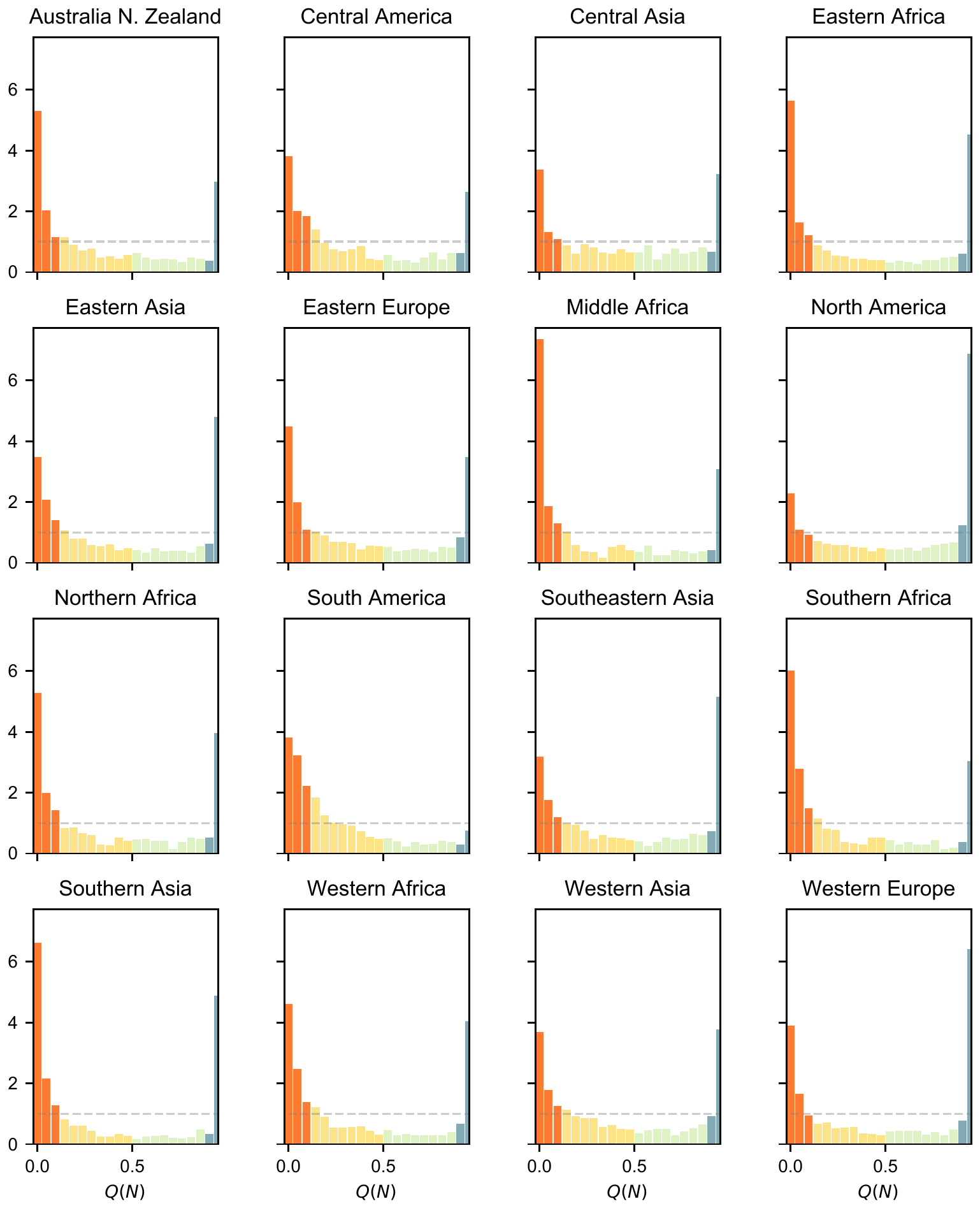}
\caption{Quantile of the number of BUCs in a tile according to the theoretical distribution $P(N| A^{tot}_{BUC})$. }
\label{fig:appendix:quantile_histograms}
\end{figure}

\begin{figure}[ht]
\centering
\includegraphics[width=0.9\textwidth]{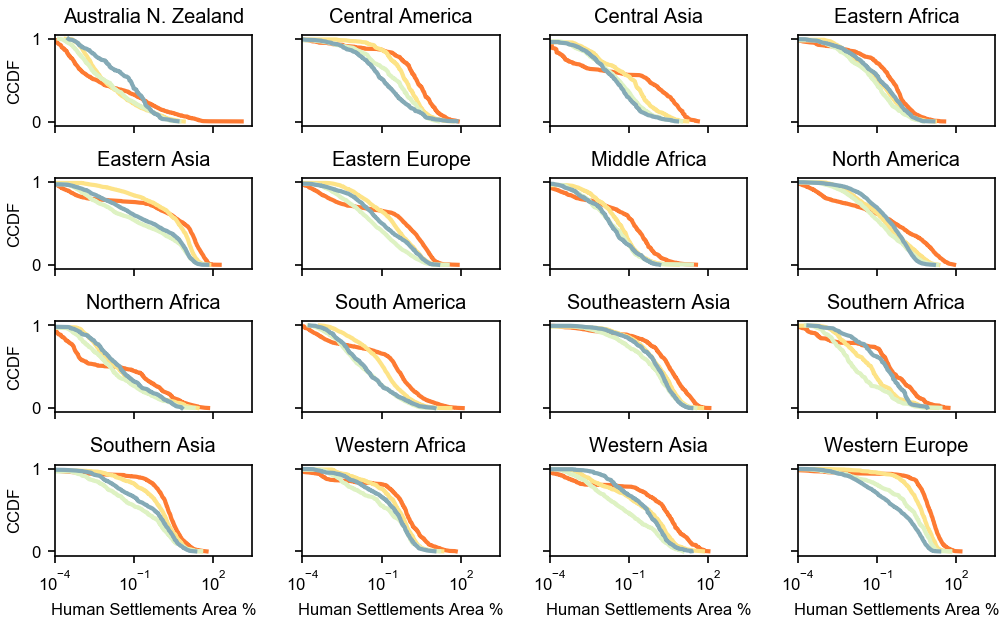}
\caption{Counter cumulative distribution function (CCDF) of the HS areas separately for each class of settlement patterns: Dispersion (blue), Balanced (green and yellow) and Agglomeration (orange). Not surprisingly, the tiles in the Agglomeration class contain a higher number of large clusters. The cluster size distributions of the tiles in the Dispersion class are not consistently different from those in the Balanced class.}
\label{fig:appendix:ccdf_builtup}
\end{figure}

\begin{figure}[ht]
\centering
\includegraphics[width=0.7\textwidth]{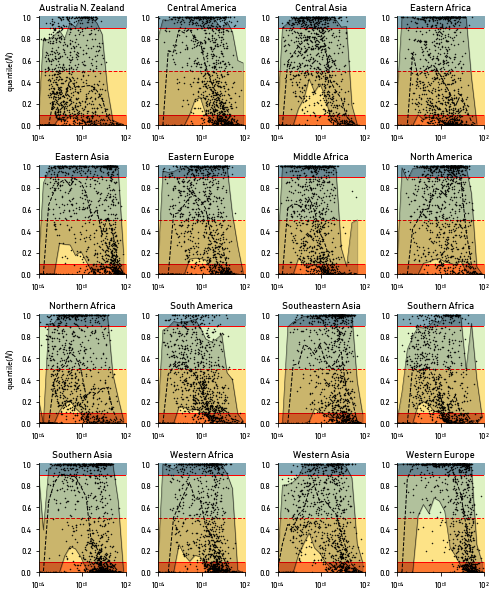}
\caption{Scatter distribution of the quantile of the number of BUCs in a tile according to the theoretical distribution $P(N| A^{tot}_{BUC})$. }
\label{fig:appendix:quantile_histograms_scatter}
\end{figure}

\clearpage
\section*{Robustness tests}
\begin{figure}[ht]
\centering
\includegraphics[width=\textwidth]{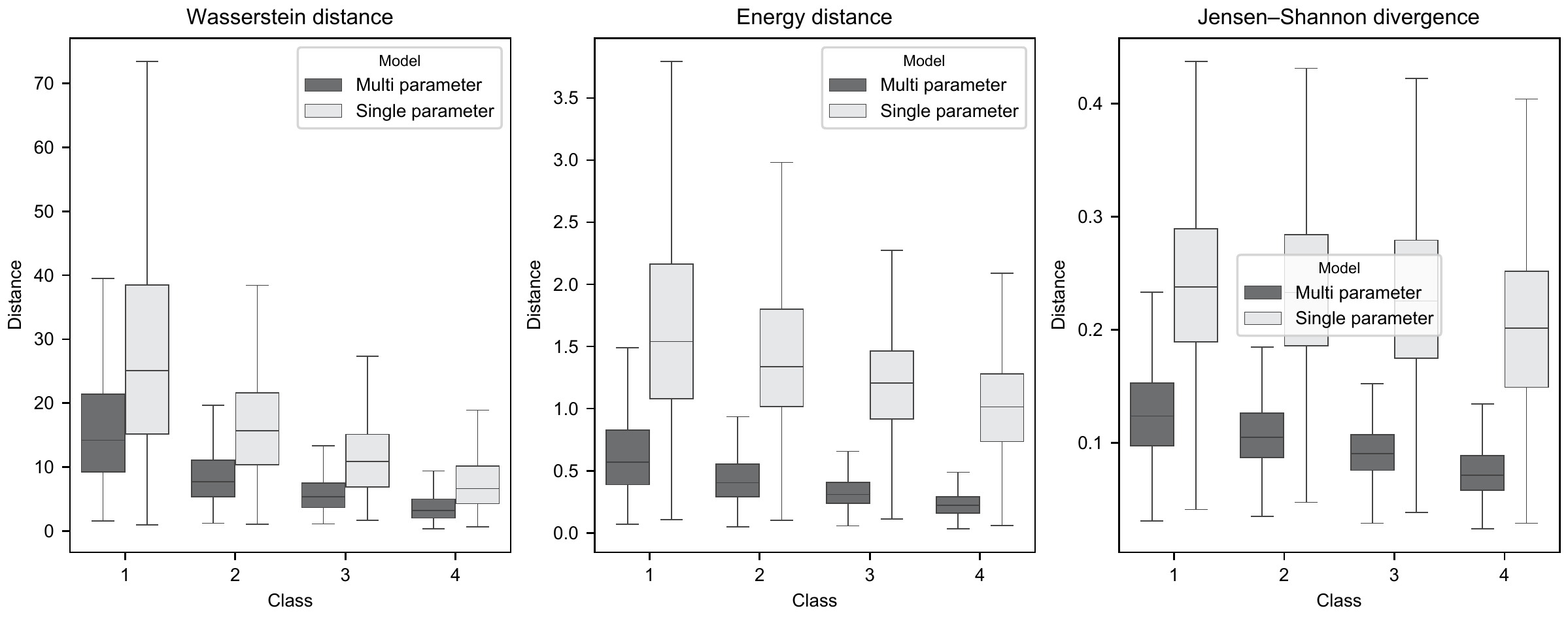}
\caption{Distance between each real tile and its best simulation for the multi-parameter and single-parameter models. Our model consistently achieves shorter distances between simulated and real tiles, even when using alternative metrics such as the Energy distance and the Jensen–Shannon divergence.}
\label{fig:appendix:distances}
\end{figure}

We run several tests to verify the sensitivity of the predictive model to:
\begin{itemize}
    \item \emph{Alternative metrics for the matches.} The match between real and simulated tiles might be sensitive to the choice of the similarity metric. Therefore, in \Cref{fig:appendix:distances}, we show the results with the energy distance~\cite{energy2013} (i.e. Wasserstein-2 distance), the Jensen-Shannon (JS) divergence, and the Wasserstein distance (i.e. Earth mover's distance). \Cref{table:tradeoff} shows the same result broken down per class.
    The energy distance is defined as the distance between two independent random variables $X$, $Y$ as:
\begin{equation}
    D^2(X, Y) = 2 \mathbb{E} \left| X - Y \right| - \mathbb{E} \left| X - X' \right| - \mathbb{E} \left| Y - Y' \right| 
\end{equation}
where $\mathbb{E} \left|X\right|<\infty$, $\mathbb{E} \left|Y\right|<\infty$, $X'$ is an iid copy of $X$ and $Y'$ is an iid copy of $Y$.
     The JS divergence is instead defined as:
    \begin{equation*}
        D_\text{JS}(P \parallel Q)= \frac{1}{2}D_\text{KL}(P \parallel M)+\frac{1}{2}D_\text{KL}(Q \parallel M)
    \end{equation*}
    where $P$ and $Q$ are discrete probability and $M=\frac{1}{2}(P+Q)$.
    
    \item \emph{The multi-prob parameter model.} We also test for a different formalization of the multi-parameter model where the exponent $\gamma$ is not changed only one time but is instead chosen at random with a specified probability in each stage of the simulation process.
    We simulate the growth in urban area through a two-dimensional $N \times N$ lattice whose sites $w_{i,j}$ can be either occupied ($w_{i,j}=1$) or empty ($w_{i,j}=0$). Without loss of generality, we set the initial configuration with $w_{N/2, N/2} = 1$ and all other pixels are zeros. Then, we simulate an evolution process where in each step, the probability that each empty site will be occupied is:
    \begin{equation*}
        q_{i,j} = C \frac{\sum_{k}^N\sum_{z}^N w_{k,z} d_{k,z}^{-\Gamma}}{\sum_{k}^N\sum_{z}^N d_{k,z}^{-\Gamma}}
    \end{equation*}
    where $C=1/max_{i,j}(q_{i,j})$ is a normalization constant for each step and $d_{k,z}$ is the Euclidean distance between site $w_{i,j}$ and site $w_{k,z}$. $\Gamma$ is chosen based on a number $p$ that is randomly chosen in each step:
  \begin{equation*}
    \Gamma=
    \begin{cases}
      \gamma_1, & \text{if}\ p<s \\
      \gamma_2, & \text{otherwise}
    \end{cases}
  \end{equation*}
  where $s$ is a chosen probability threshold of the simulation and $\gamma_1, \gamma_2$ are selected growth parameters of the simulation. In each step, $w_{i,j}=1$ iff $q_{i,j}>0.5$. We stop the growth of urban areas when $\frac{1}{N^2} \sum_{i,j}^N w_{i,j} \geq 0.5$.
    Since we choose the $\Gamma$ parameter in each step, we call this the multi-prob parameter model, whereas the other one is called multi-parameter model.
    \Cref{table:parameters} shows the simulated parameters. \Cref{table:f1classes} shows that the alternative formulation has comparable results of the presented mode, in terms of the F1-score between classes.
\end{itemize}
Together, these results confirm the robustness of our models and methods.

\begin{table}[ht]
    \ra{1.2}
    \centering
    \caption{2-way Kolmogorov-Smirnov test of the empirical distribution of the distances between real tiles and simulated tiles for the various models and the single-parameter model. The multi parameter model we propose achieve better performance than the single-parameter model in all classes. (**) indicates a $p$-value $< 0.001$.}
    \begin{tabular}{@{}lrrrrr@{}}
        \toprule
        \textbf{Method} &  \textbf{Agglomeration} & \multicolumn{2}{c}{\textbf{Balanced}} & \textbf{Dispersion} & \textbf{All}
		\\
		& \cellcolor{agglomerationColor} & \cellcolor{balanced1Color}  &  \cellcolor{balanced2Color} & \cellcolor{dispersionColor}
		\\
        \midrule
         \textbf{Multi parameter} & $0.34^{**}$ & $0.47^{**}$ & $0.48^{**}$ & $0.41^{**}$ & $0.29^{**}$  \\
         Multi-prob parameter  & $0.29^{**}$ & $0.38^{**}$ & $0.39^{**}$ & $0.37^{**}$ & $0.26^{**}$\\
         \bottomrule
    \end{tabular}
    \label{table:tradeoff}
\end{table}

\begin{table*}[tbhp!]
\setlength\tabcolsep{6pt}
    \ra{1.2}
	\centering
	\caption{F1-score between the urbanization class of the real tile and the urbanization class of its most similar simulation tile. The multi-parameter model achieves the best performance in all classes.}
	\begin{tabular}{@{}lrrrrr@{}}
	    \toprule
		\textbf{Method} &  \textbf{Agglomeration} & \multicolumn{2}{c}{\textbf{Balanced}} & \textbf{Dispersion} & \textbf{All}
		\\
		& \cellcolor{agglomerationColor} & \cellcolor{balanced1Color}  &  \cellcolor{balanced2Color} & \cellcolor{dispersionColor}
		\\
		\midrule
		Single parameter & 0.66 & 0.39 & 0.24 & 0.99 & 0.48
		\\
		\textbf{Multi parameter } & 0.88 & 0.74 & 0.56 & 0.98 & 0.73
		\\
		Multi-prob parameter  & 0.89 & 0.76 & 0.61 & 0.91 & 0.74
		\\
		\bottomrule
	\end{tabular}
	
	\label{table:f1classes}
\end{table*}

\section*{Simulated parameters}
In \Cref{table:parameters} we show all the tested parameters (and their combinations) for the Multi-parameter and the Multi-prob parameter models.

\begin{table*}[tbhp!]
\setlength\tabcolsep{6pt}
	\caption{Parameters used to perform the simulations. The simulations are created from the Cartesian product of these parameters. In the multi-prob parameter model for $s=0.5$, we computed only those combinations where $y_1 < y_2$, as the probability to choose one gamma is $0.5$.}
	\centering
	\ra{1.2}
	\begin{tabular}{@{}ll@{}}
	    \toprule
		 \textbf{Parameters} &  \textbf{Values}
		\\
		\midrule
		\textbf{Single parameter model}
		\\
		$\gamma_1$ 		& all parameters with $0.002$ step from $1$ to $10$
		\\
		\midrule
		\textbf{Multi parameter model}
		\\
		$\gamma_1$ 		& $\{1, 1.4, 1.8, 2, 2.2, 2.4, 2.6, 2.8, 3, 3.2, 3.4, 3.6, 3.8, 4, 5, 6, 7, 8, 10\}$
		\\
		$\gamma_2$ 		& $\{1, 1.4, 1.8, 2, 2.2, 2.4, 2.6, 2.8, 3, 3.2, 3.4, 3.6, 3.8, 4, 5, 6, 7, 8, 10\}$
		\\
		$s$ 		& \specialcell{$\{.0002, .00005, .0008, .0001, .0004, .0006,$\\$.001, .002, .004, .006, .008, .01, .02, .03, .04, .05, .06, .07, .08, .09,$\\$ .1, .2, .3, .4, .5\}$}
		\\
		\midrule
		\textbf{Multi-prob parameter model}
		\\
		$\gamma_1$ 		& $\{1, 1.4, 1.8, 2, 2.2, 2.4, 2.6, 2.8, 3, 3.2, 3.4, 3.6, 3.8, 4, 5, 6, 7, 8, 10\}$
		\\
		$\gamma_2$ 		& $\{1, 1.4, 1.8, 2, 2.2, 2.4, 2.6, 2.8, 3, 3.2, 3.4, 3.6, 3.8, 4, 5, 6, 7, 8, 10\}$
		\\
		$s$ 		& \specialcell{$\{0.5, 0.51, 0.52, 0.54, 0.57, 0.59, 0.61,$\\$ 0.64, 0.66, 0.68, 0.71, 0.73, 0.75, 0.77, 0.79, 0.82, 0.84, 0.86, 0.89,$\\$ 0.91,
             0.93, 0.96, 0.98, 0.99\}$}
		\\
		\bottomrule
	\end{tabular}
	
	\label{table:parameters}
\end{table*}

\section*{Regressions with urban indicators}
Here, we provide some additional details for the regressions we mention in the main paper.

The regressions are fitted with the Ordinary Least Squares (OLS) regression. We computed all the urban indicators per each tile by intersecting the tiles grid with the data. 

We test two models, namely M1 and M2. The former uses only the HSs area as a independent variable, while the latter uses the HSs area and the quantile (of the tile) obtained from the devations from the scaling theory.
In \Cref{tab:regressions} we show the $\beta$ coefficients and the $R^2$ for all the models we tested, while in \Cref{tab:regressions2} we show the rank regression, which is non-linear.

\begin{table}[!h]
\footnotesize
    \centering
    \caption{OLS regression between variables. The numbers filled in the table are the $\beta$ coefficient of the OLS regression. The last row corresponds to the $R^2$ coefficient, which represents the goodness of fit.}
    \label{tab:regressions}
    \begin{tabular}{@{}l rr rr rr rr rr@{}}
    \midrule
    \textbf{Variables} &
    \multicolumn{2}{c}{\textbf{on-road CO2}} &  \multicolumn{2}{c}{\textbf{on-road CO2 per capita}} & 
    \multicolumn{2}{c}{\textbf{Roads per capita}} & 
    \multicolumn{2}{c}{\textbf{Car commuting}} & 
    \multicolumn{2}{c}{\textbf{Public transportation}}\\
    \cmidrule(l{2pt}){2-3} \cmidrule(l{2pt}){4-5} \cmidrule(l{2pt}){6-7} \cmidrule(l{2pt}){8-9} \cmidrule(l{2pt}){10-11}
    & M1 & M2 & M1 & M2 & M1 & M2 & M1 & M2 & M1 & M2\\
    \midrule
    HSs area & -458046 & 568544 & -42.18 & -26.81 & -0.004 & -0.003 & -0.013 & -0.010 & 0.015 & 0.015\\
    $Q(N)$ & - & 2382267 & - & 35.66 & - & 0.002 & - & 0.009 & - & -0.001\\
    \midrule
    $R^2$ & 0.01 & \textbf{0.16} & 0.31 & \textbf{0.50} & 0.42 & \textbf{0.49} & 0.06 & \textbf{0.09} & 0.23 & \textbf{0.24} \\
    \bottomrule      
    \end{tabular}
\end{table}

\begin{table}[!h]
\footnotesize
    \centering
    \caption{OLS regression between ranks. The numbers filled in the table are the $\beta$ coefficient of the OLS regression. The last row corresponds to the $R^2$ coefficient, which represents the goodness of fit.}
    \label{tab:regressions2}
    \begin{tabular}{@{}l rr rr rr rr rr@{}}
    \midrule
    \textbf{Variables} &
    \multicolumn{2}{c}{\textbf{on-road CO2}} &  \multicolumn{2}{c}{\textbf{on-road CO2 per capita}} & 
    \multicolumn{2}{c}{\textbf{Roads per capita}} & 
    \multicolumn{2}{c}{\textbf{Car commuting}} & 
    \multicolumn{2}{c}{\textbf{Public transportation}}\\
    \cmidrule(l{2pt}){2-3} \cmidrule(l{2pt}){4-5} \cmidrule(l{2pt}){6-7} \cmidrule(l{2pt}){8-9} \cmidrule(l{2pt}){10-11}
    & M1 & M2 & M1 & M2 & M1 & M2 & M1 & M2 & M1 & M2\\
    \midrule
    HSs area & -24.91 & 41.72 & -148.62 & -106.31 & -160.53 & -146.13 & -7.81 & 31.49 & 78.01 & 46.07\\
    $Q(N)$ & - & 108.18 & - & 68.68 & - & 23.46 & - & 83.18 & - & -67.57\\
    \midrule
    $R^2$ & 0.02 & \textbf{0.24} & 0.67 & \textbf{0.76} & 0.76 & \textbf{0.77} & 0.00 & \textbf{0.13} & 0.14 & \textbf{0.23} \\
    \bottomrule      
    \end{tabular}
\end{table}

\clearpage
\section*{Global HSs Density}
In \Cref{table:bins} we show the exact number of the density of HSs in all the macro-areas.

\begin{table}[ht!]
    \ra{1.2}
    \centering
    \caption{Cumulative probability of HS areas for all the areas around the globe.}
    \begin{tabular}{@{}lrrrrrr@{}}
        \toprule
        \textbf{Method} &  \multicolumn{6}{c}{\textbf{Lower bounds of the bins}}
		\\
		& \cellcolor{bin1} & \cellcolor{bin2}  &  \cellcolor{bin3} & \cellcolor{bin4}  & \cellcolor{bin5} & \cellcolor{bin6}
		\\
        \midrule
        Australia N. Zealand  & $0.761$ & $0.978$ & $0.991$ & $0.995$ & $0.998$ & $0.999$ \\
        Eastern Africa  & $0.246$ & $0.910$ & $0.980$ & $0.996$ & $0.999$ & $1.000$ \\
        Middle Africa  & $0.481$ & $0.983$ & $0.994$ & $0.998$ & $0.999$ & $0.999$ \\
        Northern Africa  & $0.739$ & $0.958$ & $0.982$ & $0.996$ & $0.999$ & $0.999$ \\
        Southern Africa & $0.386$ & $0.911$ & $0.964$ & $0.993$ & $0.998$ & $0.999$ \\
        Western Africa & $0.415$ & $0.873$ & $0.963$ & $0.991$ & $0.999$ & $0.999$ \\
        Central America  & $0.145$ & $0.732$ & $0.890$ & $0.966$ & $0.995$ & $0.999$ \\
        North America  & $0.491$ & $0.855$ & $0.937$ & $0.975$ & $0.995$ & $0.999$ \\
        South America  & $0.453$ & $0.946$ & $0.982$ & $0.994$ & $0.998$ & $0.999$ \\
        Central Asia  & $0.537$ & $0.928$ & $0.961$ & $0.986$ & $0.998$ & $0.999$ \\
        Eastern Asia  & $0.454$ & $0.673$ & $0.741$ & $0.832$ & $0.970$ & $0.995$ \\
        Southeastern Asia & $0.120$ & $0.540$ & $0.752$ & $0.910$ & $0.988$ & $0.998$ \\
        Southern Asia & $0.156$ & $0.550$ & $0.827$ & $0.963$ & $0.998$ & $0.999$ \\
        Western Asia & $0.427$ & $0.820$ & $0.936$ & $0.979$ & $0.997$ & $0.999$ \\
        Eastern Europe  & $0.577$ & $0.857$ & $0.941$ & $0.991$ & $0.999$ & $0.999$ \\
        Western Europe & $0.095$ & $0.385$ & $0.572$ & $0.833$ & $0.990$ & $0.999$ \\
         \bottomrule
    \end{tabular}
    \label{table:bins}
\end{table}

\clearpage
\section*{Global classification of the HSs patterns}
\begin{figure}[ht]
\centering
\includegraphics[width=0.6\textwidth]{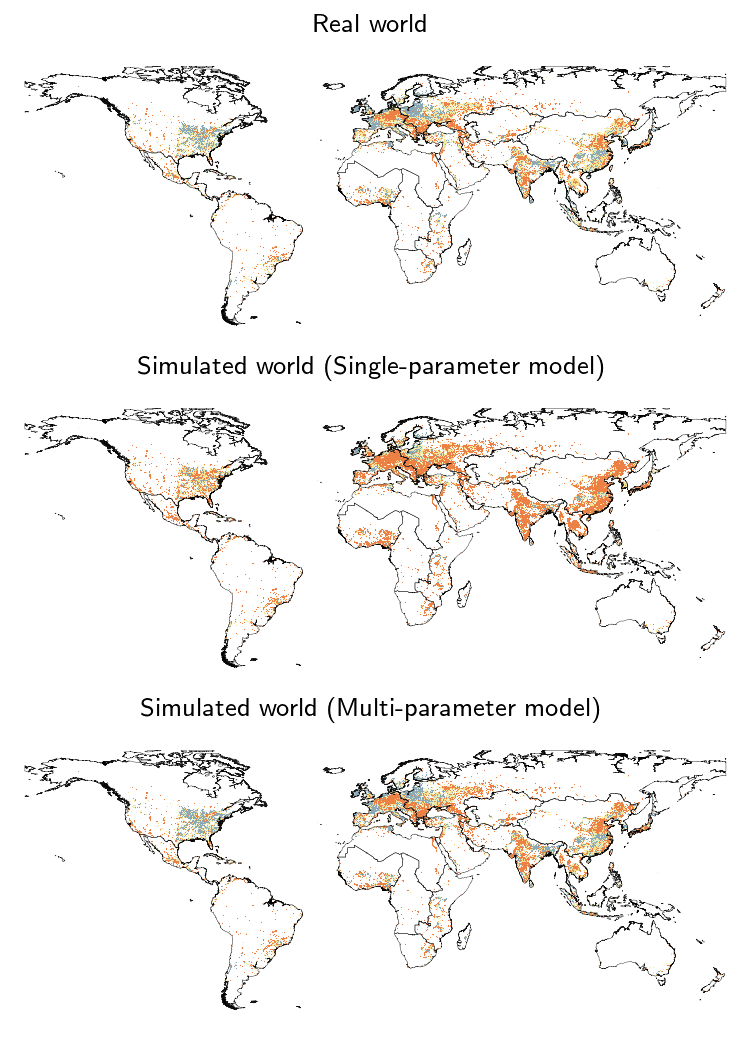}
\caption{Global classification of all the tiles with more than 1\% urbanization for the real tiles (first row) and the simulated ones with the Single-parameter model (second row) and the Multi-parameter model (third row). As it can be seen, the Single-parameter model overestimates the number of tiles in the Dispersion and Agglomeration classes. The multi-parameter model mitigates this problem and it is very similar to the real classification of the tiles (as shown by the F1-score accuracy over the different classes). This means that the Multi-parameter model reliably simulates the global pattern of urbanization.}
\label{fig:appendix:simulated-world}
\end{figure}

In \Cref{fig:appendix:simulated-world} we show the classification of all the tiles with more than 1\% urbanization. In this figure we compare the real tiles with the simulated ones.

\clearpage
\section*{Some examples of simulations}

\begin{figure}[ht]
\centering
\includegraphics[width=0.8\textwidth]{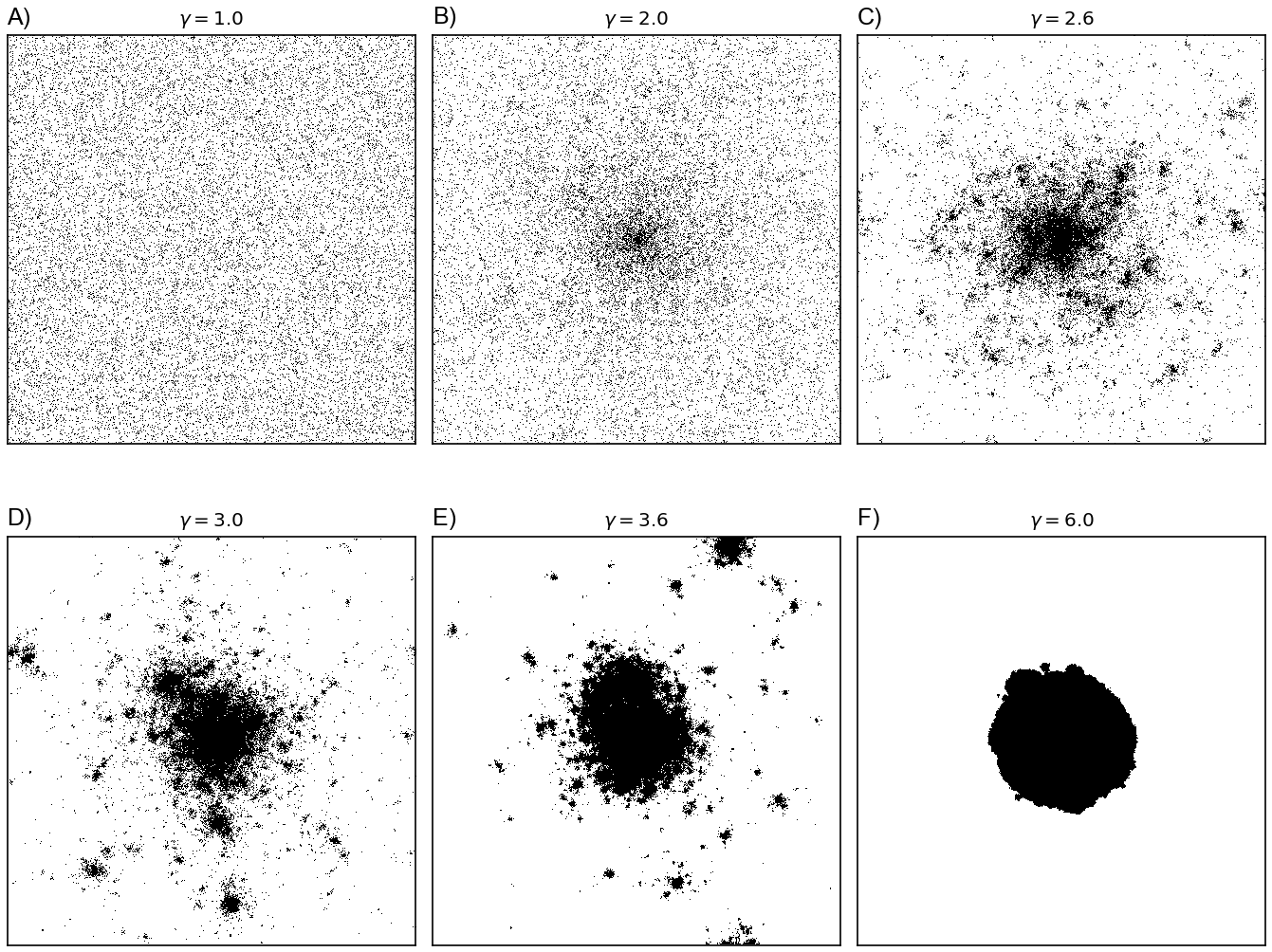}
\caption{Different simulations of the single-parameter model with 30\% of total BUC at different $\gamma$ values. Low values of $\gamma$ generate dispersed settlement patterns, whereas high values of $\gamma$ generate compact patterns. It can be seen that in A) the model generates a random noise pattern, as the urban areas are created without caring on the existing urban areas ($\gamma = 0.1$). Contrarily, in F) a dense urban pattern is generated, as the model creates new urban ares with high probability only near those areas that are already built-up.}
\label{fig:appendix:explain_rybski}
\end{figure}

\begin{figure}[ht]
\centering
\includegraphics[width=0.8\textwidth]{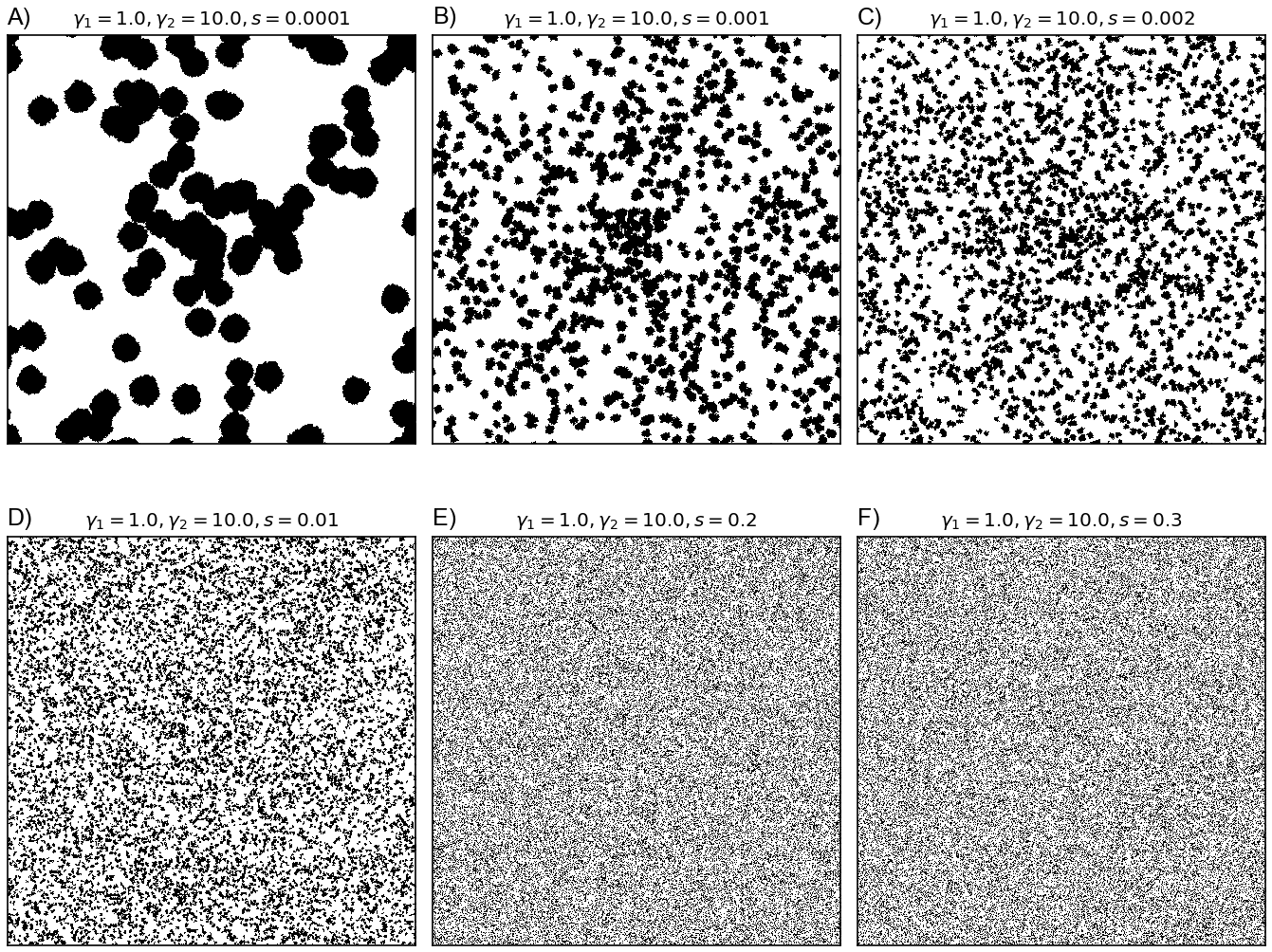}
\caption{Different simulations of the Multi-parameter model with 30\% of total BUC for the same $\gamma_1$ and $\gamma_2$ but different $s$ values. It can be seen that in A) the model starts with a sparse pattern ($\gamma_1 = 0.1$) and then switches to the dense one ($\gamma_1 = 10.0$) until 30\% of urbanization. The resulting pattern is clustered in circles. Contrarily, in F) a random noise pattern is created, as the urban areas are created without caring on the existing urban areas ($\gamma_1 = 0.1$). The parameter $\gamma_1 = 0.1$ is kept until the end of the simulation. The Multi-parameter model generates more complex patterns than the Single-parameter model.}
\label{fig:appendix:explain_multi}
\end{figure}

\begin{figure}[ht]
\centering
\includegraphics[width=0.8\textwidth]{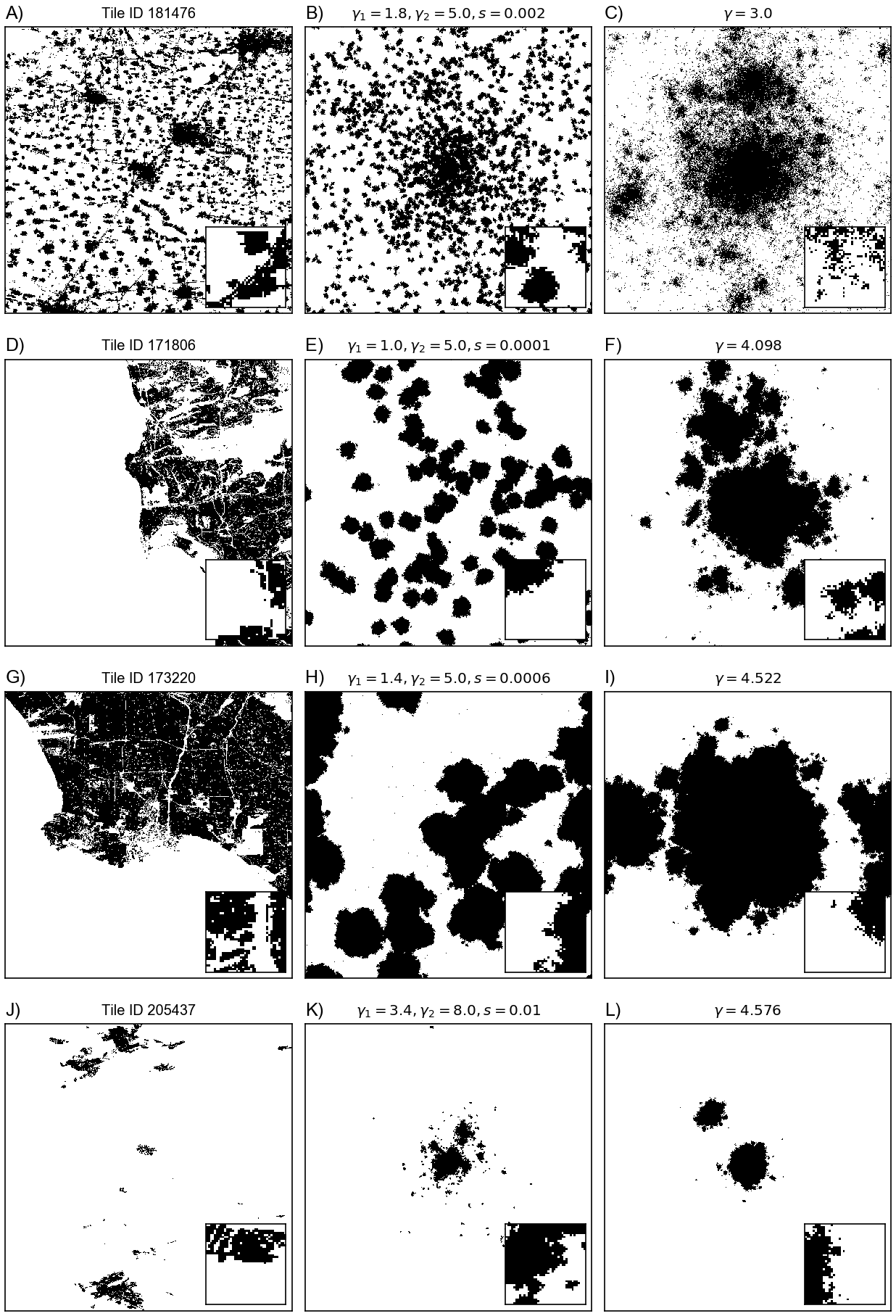}
\caption{Some examples of the best simulations for four tiles in the Agglomeration class. The left column shows the real tile, the central column shows the most similar tile generated with the multi-parameter model, and the right column shows the single-parameter model simulation that is most similar to the real tile.}
\label{fig:appendix:class1}
\end{figure}
\begin{figure}[ht]
\centering
\includegraphics[width=0.8\textwidth]{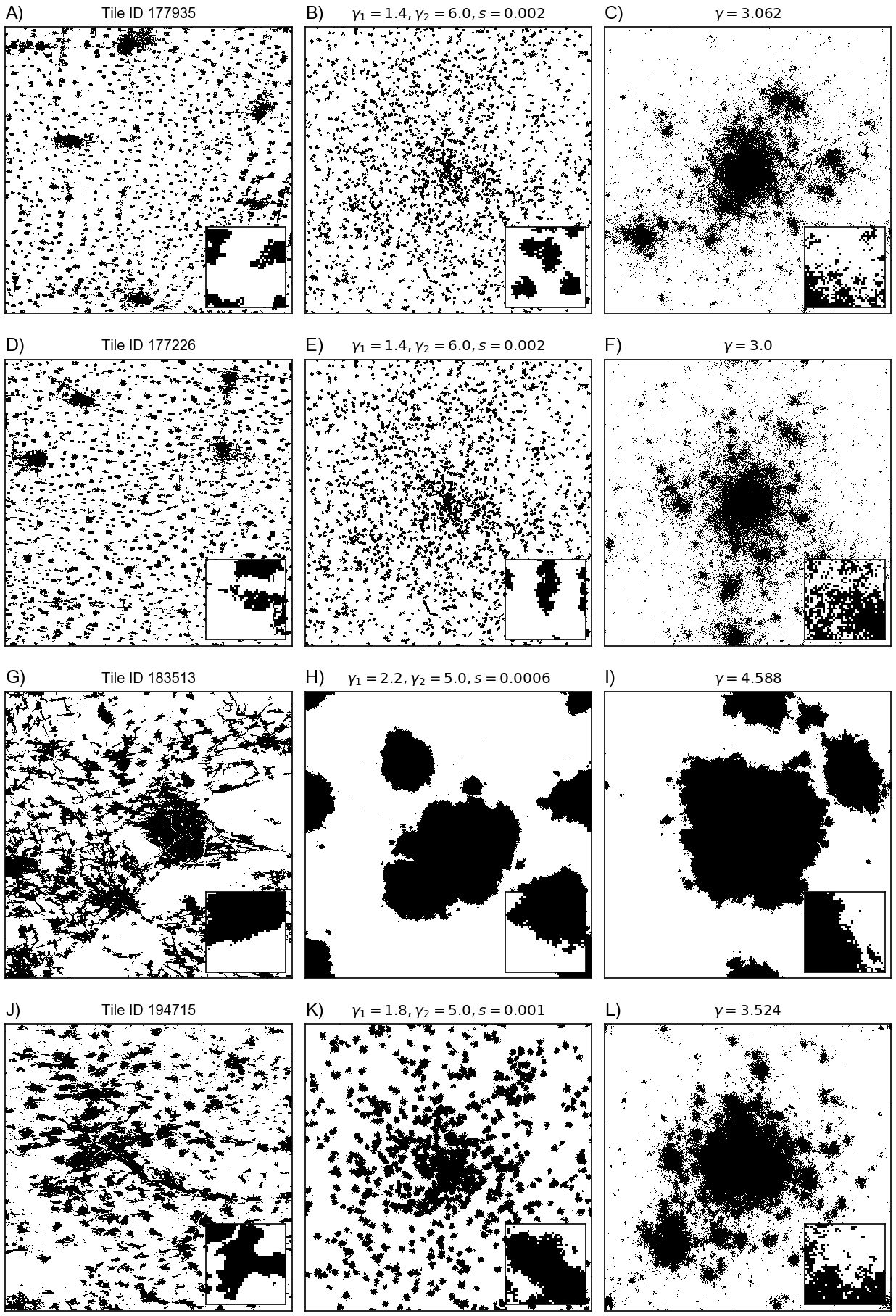}
\caption{Some more examples of the best simulations for four tiles in the Agglomeration class. The left column shows the real tile, the central column shows the most similar tile generated with the multi-parameter model, and the right column shows the single-parameter model simulation that is most similar to the real tile.}
\label{fig:appendix:class11}
\end{figure}

\begin{figure}[ht]
\centering
\includegraphics[width=0.8\textwidth]{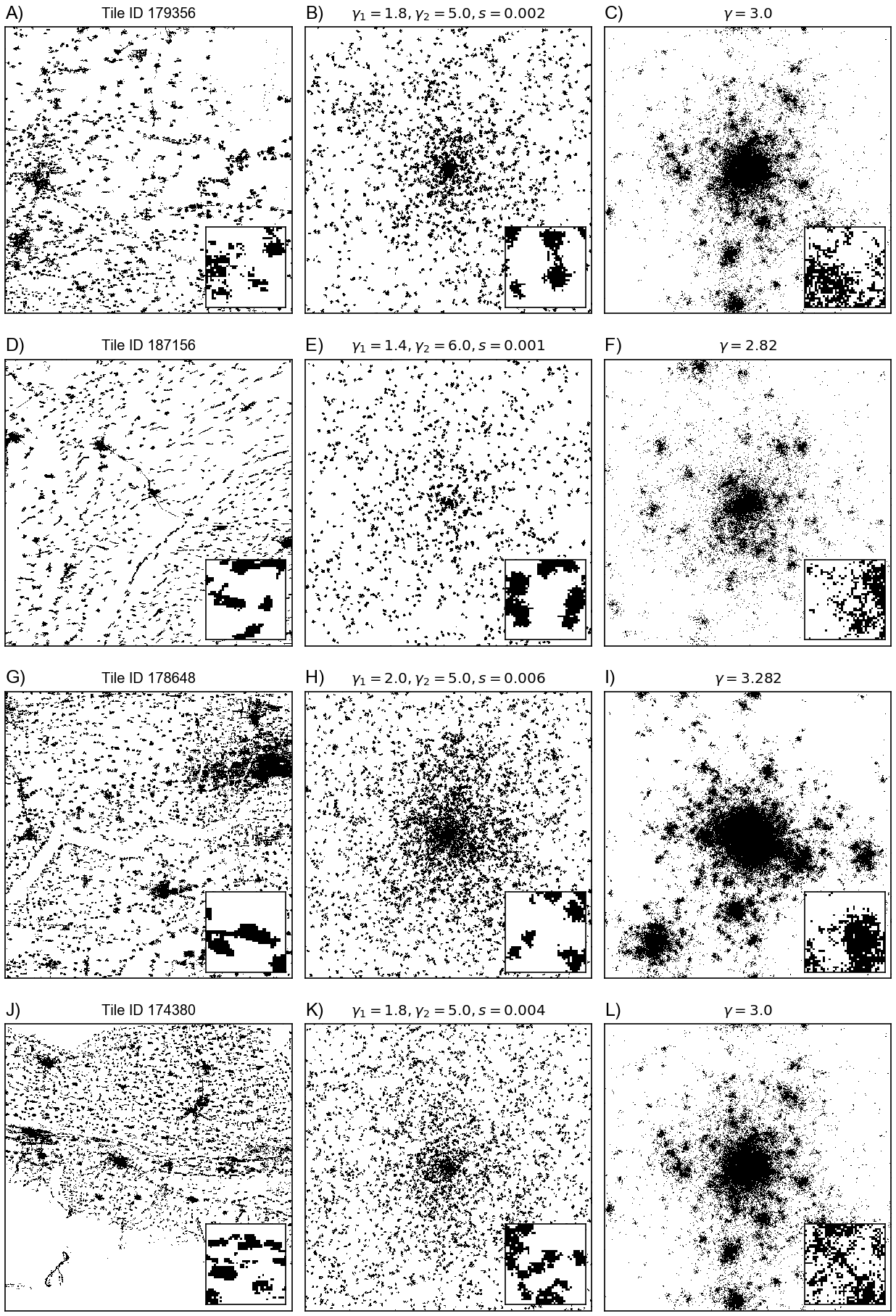}
\caption{Some examples of the best simulations for four tiles in the Balanced class (yellow group). The left column shows the real tile, the central column shows the most similar tile generated with the multi-parameter model, and the right column shows the single-parameter model simulation that is most similar to the real tile.}
\label{fig:appendix:class2}
\end{figure}
\begin{figure}[ht]
\centering
\includegraphics[width=0.8\textwidth]{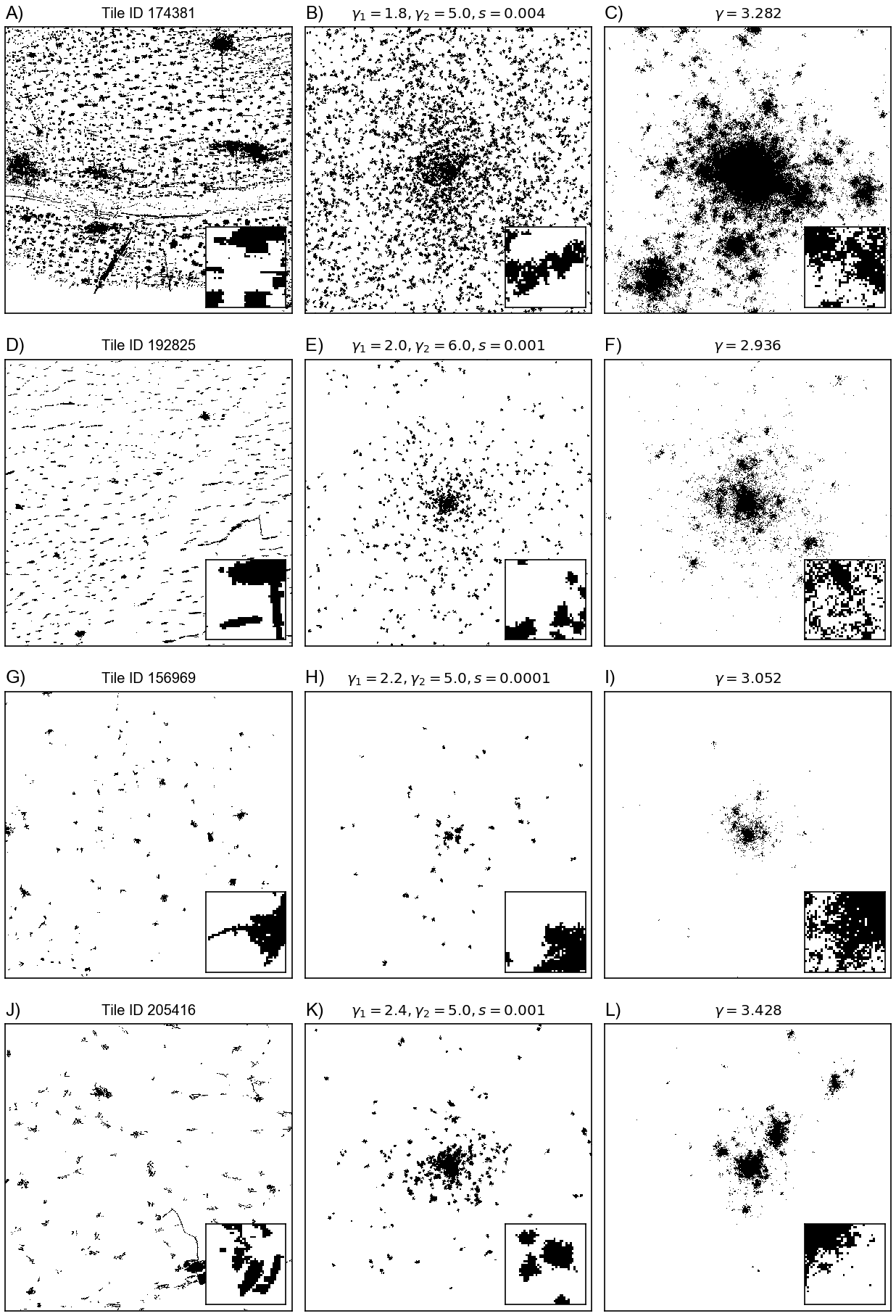}
\caption{Some more examples of the best simulations for four tiles in the Balanced class (yellow group). The left column shows the real tile, the central column shows the most similar tile generated with the multi-parameter model, and the right column shows the single-parameter model simulation that is most similar to the real tile.}
\label{fig:appendix:class21}
\end{figure}

\begin{figure}[ht]
\centering
\includegraphics[width=0.8\textwidth]{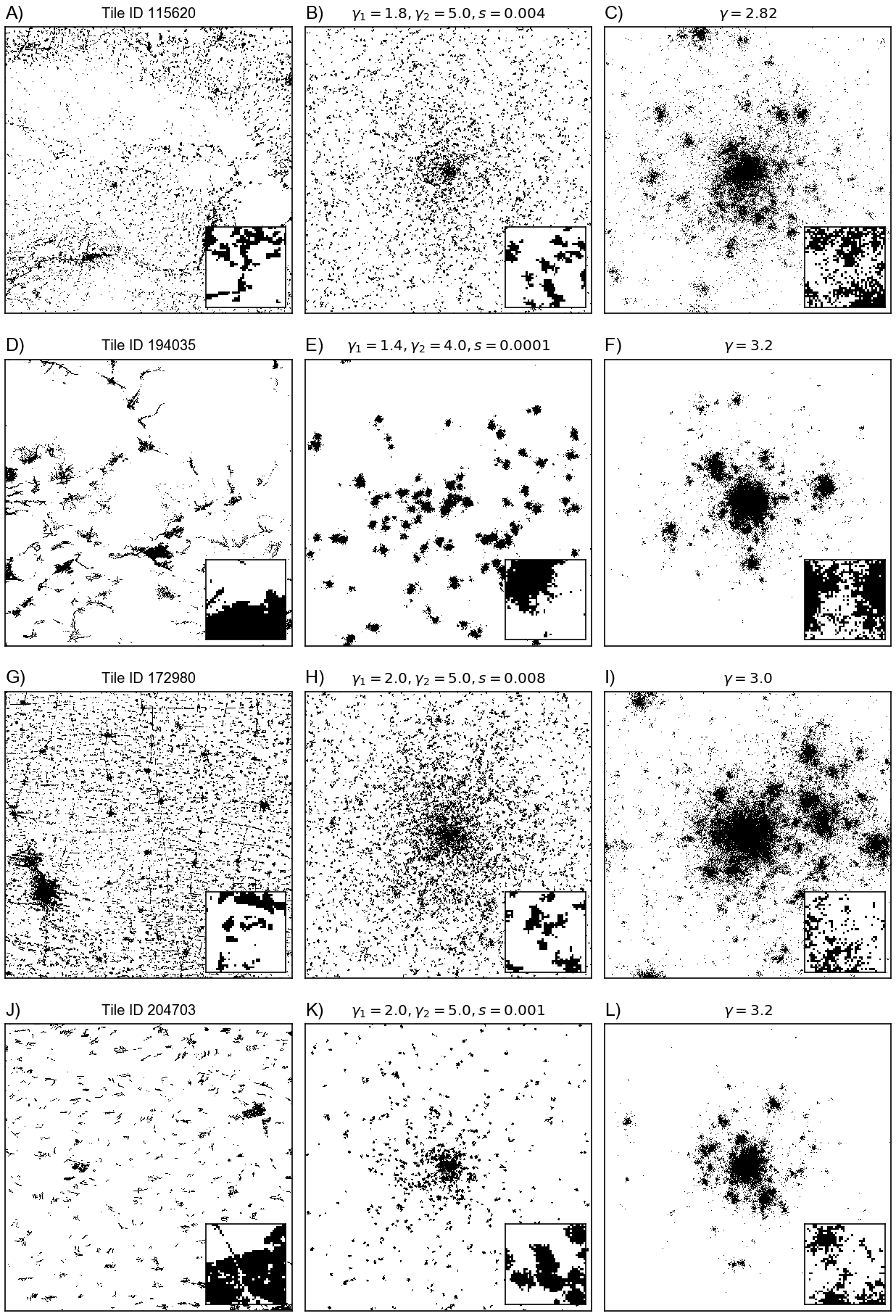}
\caption{Some examples of the best simulations for four tiles in the Balanced class (green group). The left column shows the real tile, the central column shows the most similar tile generated with the multi-parameter model, and the right column shows the single-parameter model simulation that is most similar to the real tile.}
\label{fig:appendix:class3}
\end{figure}
\begin{figure}[ht]
\centering
\includegraphics[width=0.8\textwidth]{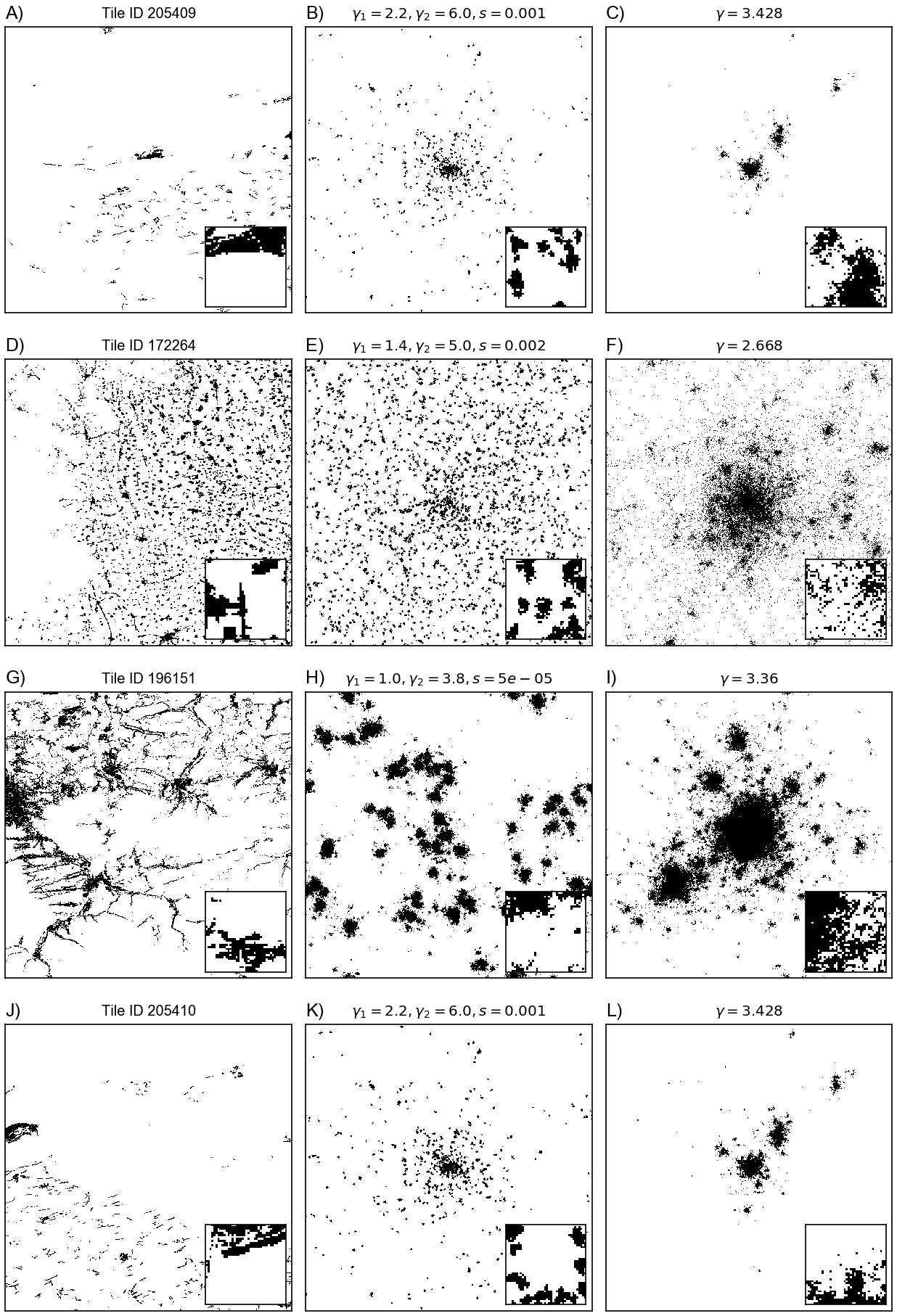}
\caption{Some more examples of the best simulations for four tiles in the Balanced class (green group). The left column shows the real tile, the central column shows the most similar tile generated with the multi-parameter model, and the right column shows the single-parameter model simulation that is most similar to the real tile.}
\label{fig:appendix:class31}
\end{figure}

\begin{figure}[ht]
\centering
\includegraphics[width=0.8\textwidth]{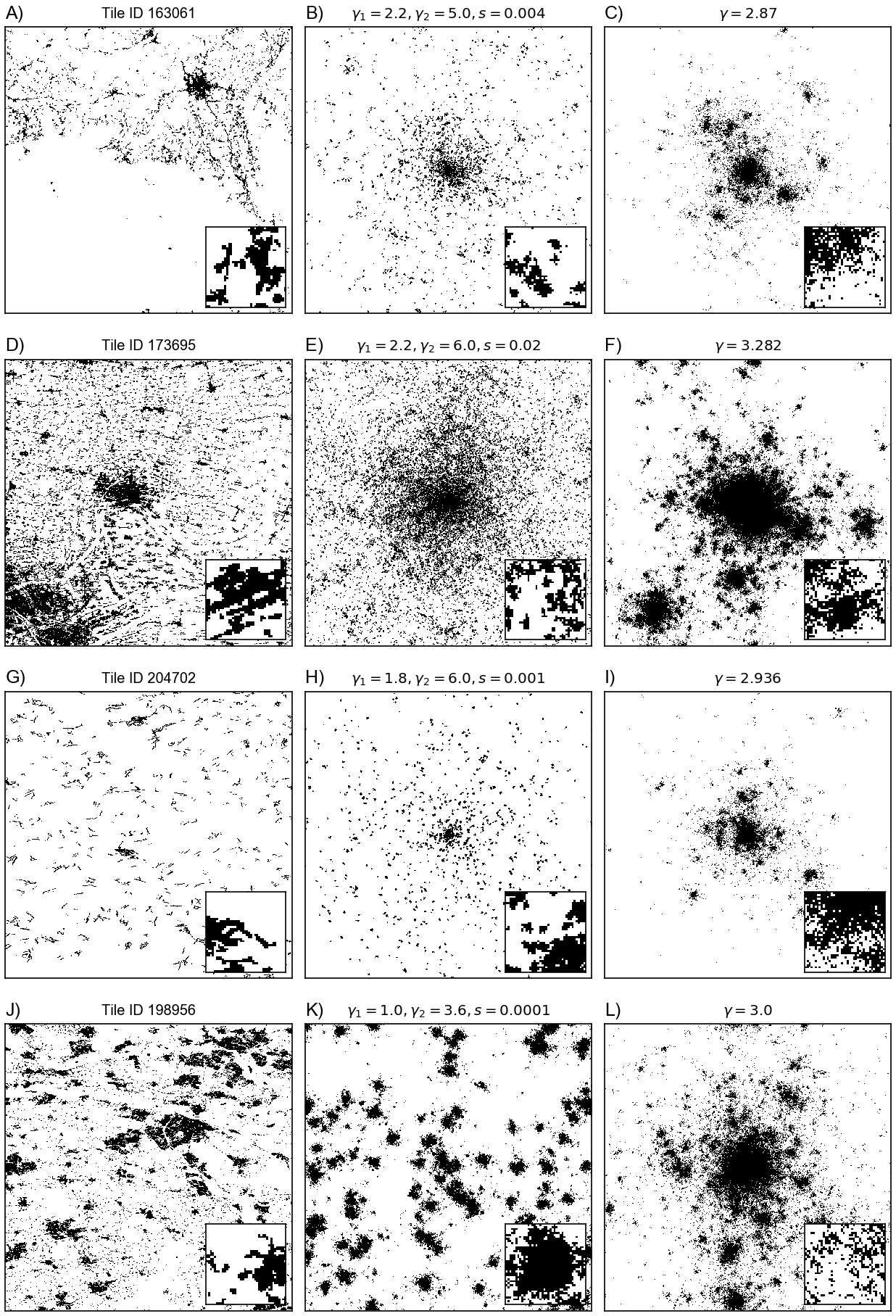}
\caption{Some examples of the best simulations for four tiles in the Dispersion class. The left column shows the real tile, the central column shows the most similar tile generated with the multi-parameter model, and the right column shows the single-parameter model simulation that is most similar to the real tile.}
\label{fig:appendix:class4}
\end{figure}
\begin{figure}[ht]
\centering
\includegraphics[width=0.8\textwidth]{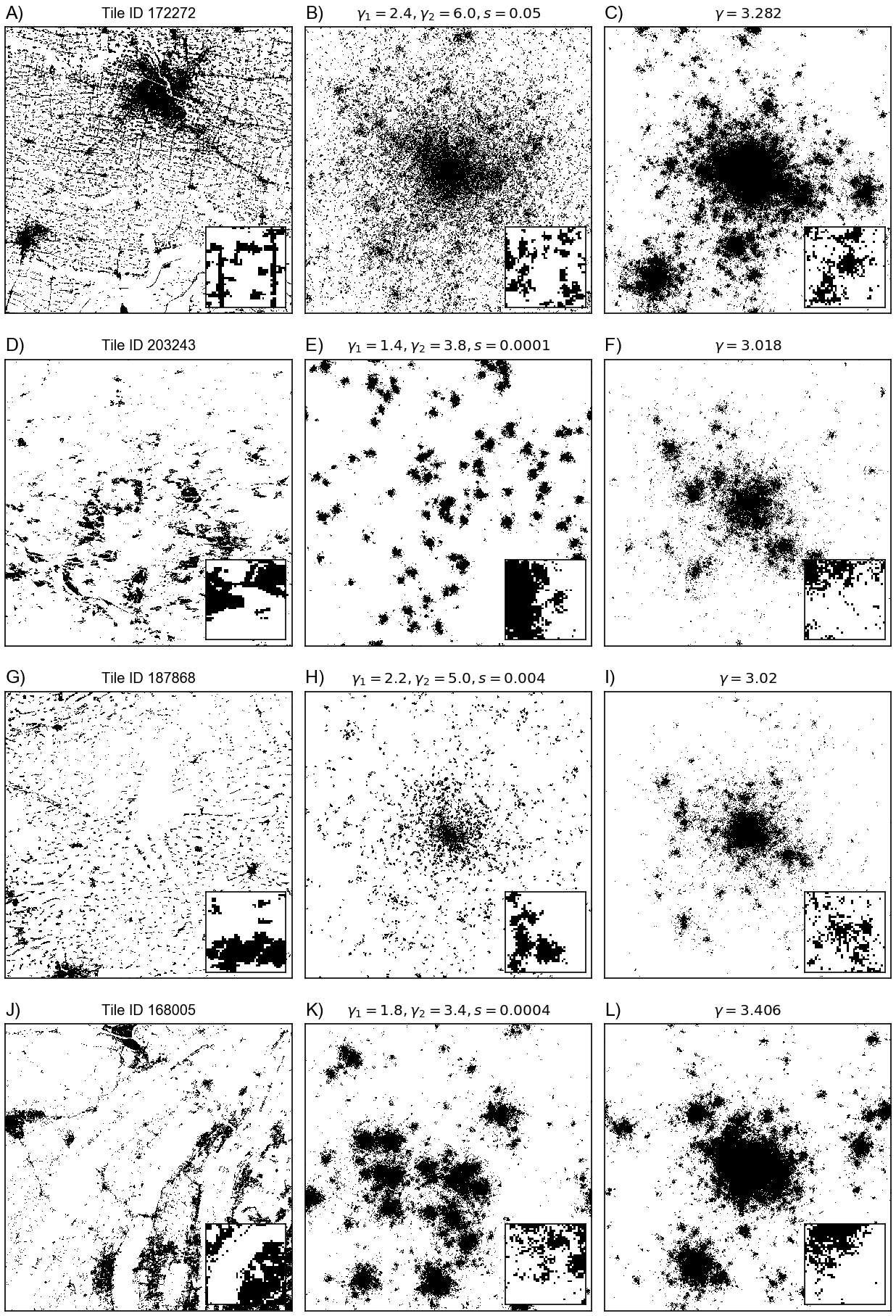}
\caption{Some more examples of the best simulations for four tiles in the Dispersion class. The left column shows the real tile, the central column shows the most similar tile generated with the multi-parameter model, and the right column shows the single-parameter model simulation that is most similar to the real tile.}
\label{fig:appendix:class41}
\end{figure}

\begin{figure}[ht]
\centering
\includegraphics[width=0.8\textwidth]{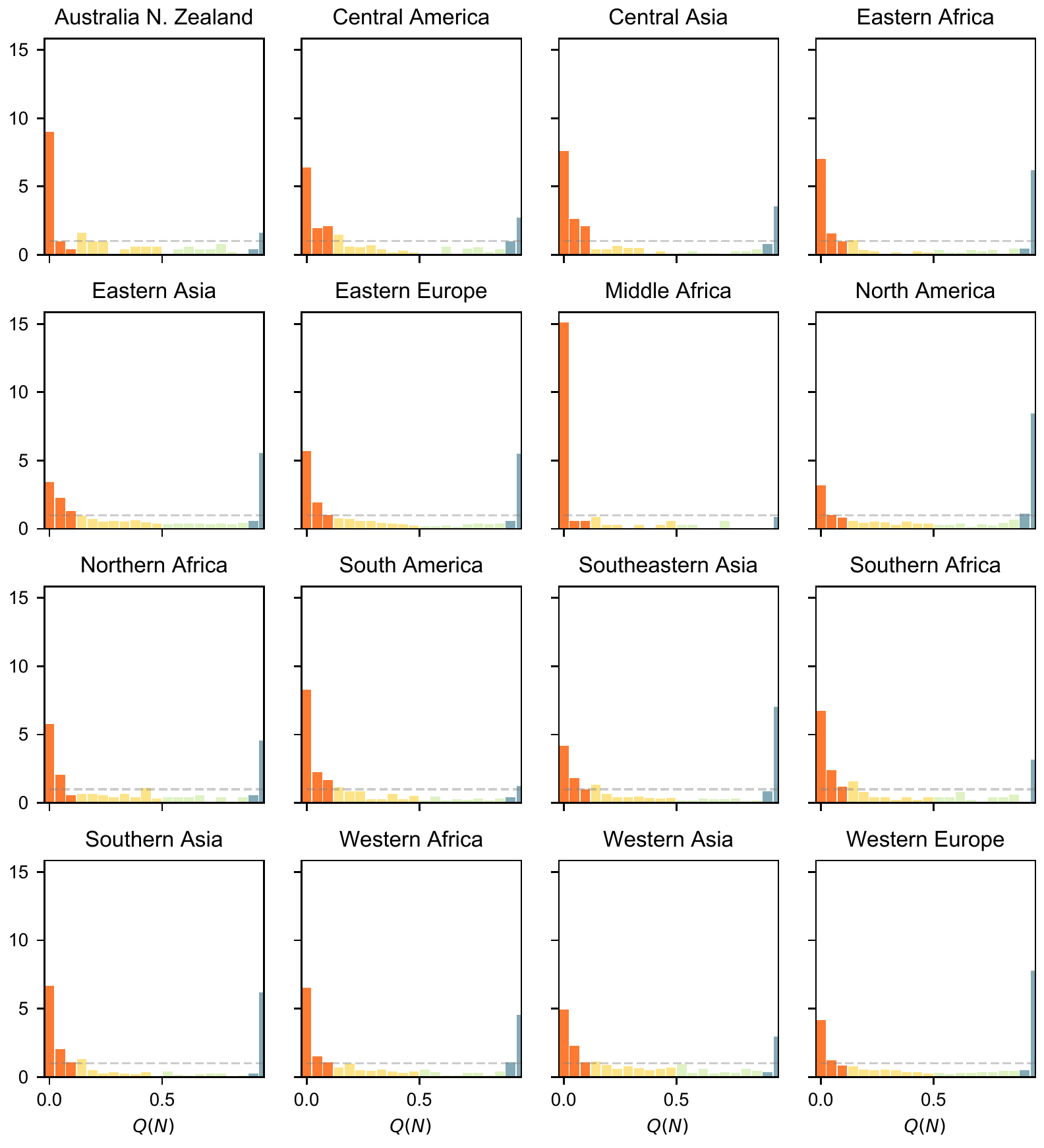}
\caption{Agglomeration Dispersion dichotomy in multi-parameter model generated tiles.}
\label{fig:appendix:dispersion-model}
\end{figure}

\end{document}